\documentclass[useAMS,usenatbib]{mn2e}
\usepackage{graphicx}
\usepackage{amsmath}
\usepackage{epstopdf}
\usepackage{grffile}
\usepackage{amssymb}
\usepackage{lscape}
\newcommand{\Om}{\Omega_{\rm m}}
\newcommand{\Ob}{\Omega_{\rm b}}
\newcommand{\OL}{\Omega_{\Lambda}}

\newcommand{\powlt}{\alpha_{\rm LT}}
\newcommand{\normlt}{C_{\rm LT}}
\newcommand{\zevollt}{\gamma_{\rm z,LT}}
\newcommand{\scattlt}{\sigma_{lnL|T}}

\newcommand{\powmt}{\alpha_{\rm MT}}
\newcommand{\normmt}{C_{\rm MT}}
\newcommand{\zevolmt}{\gamma_{\rm z,MT}}
\newcommand{\scattmt}{\sigma_{lnT|M}}

\newcommand{\crhat}{\widehat{\rm CR}}
\newcommand{\hrhat}{\widehat{\rm HR}}
\newcommand{\tot}{{\rm tot}}

\newcommand{\dndcrdhr}{dn/d{\rm CR}/d{\rm HR}}
\newcommand{\dndzdcrdhr}{dn/dz/d{\rm CR}/d{\rm HR}}
\newcommand{\dndzdm}{dn/dz/d{M}}

\newcommand{\obs}{\mathcal{O}}

\newcommand{\xc}{x_{c,0}}

\newcommand{\mdcb}{M_{\rm 200b}}
\newcommand{\mdcc}{M_{\rm 200c}}

\newcommand{\rccc}{R_{\rm 500c}}

\newcommand{\rdcb}{R_{\rm 200b}}

\title[The cosmological analysis of X-ray cluster surveys-I]{The cosmological analysis of X-ray cluster surveys: \\
I- a new method for interpreting number counts}
\author[N. Clerc et al.]{N. Clerc$^{1}$\thanks{Present e-mail: nclerc@mpe.mpg.de (MPE/Garching)} and
M. Pierre$^{1}$ and F. Pacaud$^{2}$ and T. Sadibekova$^{1}$\\
$^{1}$Laboratoire AIM, CEA/DSM/IRFU/SAp, CEA Saclay, 91191 Gif-sur-Yvette, France\\
$^{2}$Argelander-Institut f\"ur Astronomie, University of Bonn, Auf dem H\"ugel 71, 53121 Bonn, Germany}

\begin{document}

\date{Accepted 2012 April 20. Received 2012 March 28; in original form 2011 September 20}

\pagerange{\pageref{firstpage}--\pageref{lastpage}} \pubyear{2002}

\maketitle

\label{firstpage}

\begin{abstract}
We present a new method aiming to simplify the cosmological analysis of X-ray cluster surveys. It is based on purely instrumental observable quantities, considered in a two-dimensional X-ray colour-magnitude diagram (hardness ratio versus count-rate). The basic principle is that, even in rather shallow surveys, substantial information on cluster redshift and temperature is present in the raw X-ray data and can be statistically extracted; in parallel,  such diagrams can be readily predicted from an  {\em ab initio} cosmological modeling. We illustrate the methodology for the case of a 100 deg2 XMM survey having a sensitivity of $\sim 10^{-14}$~ergs/s/cm$^2$ and fit at the same time, the survey selection function, the cluster evolutionary scaling-relations and the cosmology; our sole assumption -- driven by the limited size of the sample considered in the case-study -- is that the local cluster scaling relations are known.  We devote special care to the realistic modeling of the count-rate  measurement uncertainties and evaluate the potential of the method via a Fisher analysis. In the absence of individual cluster redshifts, the CR-HR method appears to be much more efficient than the traditional approach based on cluster counts (i.e. $dn/dz$, requiring redshifts). In the case where redshifts are available, our method performs similarly as the traditional mass function ($dn/dM/dz$) for the purely cosmological parameters, but better constrains parameters defining the cluster scaling relations and their evolution. A further practical advantage of the CR-HR method is its simplicity : this fully top-down approach  totally bypasses the tedious steps consisting in deriving cluster masses from X-ray temperature measurements.

\end{abstract}


\section{Introduction}

Discriminating between different cosmic scenarios requires precision cosmological studies relying on well-controlled observables. In parallel to the Cosmological Microwave Background (CMB), Baryonic Acoustic Oscillations (BAO), Type Ia Supernovae and Weak-Lensing analyses, galaxy clusters, as the most massive bound entities in the universe, are expected to provide independent complementary constraints   \citep{Eke:1996p894,Henry:1997p5438, Oukbir:1997p4054,Borgani:2001p4880,Rozo:2007p5325,allenreview,Sehgal:2011p5963}. In particular, they appear to be  quite sensitive to the properties of the dark energy \citep{Haiman:2001p4682,Battye:2003p4789, Pierre:2011p5484}.
Cluster cosmological studies are usually based on  the cluster number counts as a function of redshift and mass. This quantity can be easily inferred from the halo model formalism and is confirmed by the most recent N-body simulations \citep{Press:1974p5302, Lacey:1993p5478,Sheth:1999p5476, Jenkins:2001p921, Springel:2005p5309} ; see \citealt{Cooray:2002p5304} for a review. Theoretically, its high sensitivity to the initial density fluctuation power-spectrum as well as its time-evolution  make it a powerful probe of structure formation. Its dependence on geometrical effects (surveyed volume) further strengthens the constraints on key cosmological quantities such as the matter content in the Universe ($\Om$) and the dark energy equation of state. From the observer's point of view, however, cluster masses are not quantities easily measurable (contrary to cluster redshifts), a fact that often leads to question the actual use of clusters as cosmological probes.

Galaxy clusters can be studied in a variety of ways, in particular through their X-ray emission. The gas trapped in the deep cluster potential is heated up to X-ray emitting temperatures. Free-free emission is the dominant mechanism from the hot plasma having a heavy element abundance of $\sim 0.3\,Z_{\odot}$;  at low temperatures ($\lesssim$2~keV) a significant fraction of the energy is emitted via recombination lines \citep{Sarazin:1988p5939}. Because extended X-ray sources at high galactic latitude almost unambiguously point toward cluster potential wells, hence  minimising projection effects, X-ray  surveys have long been considered as the ideal way of constructing cosmological cluster samples.

The {\it Einstein Observatory} Extended Medium Sensitivity Survey \citep{Gioia:1990p5474} provided the first flux-limited, X-ray selected sample of galaxy clusters, allowing pioneering cluster count analyses. It was then followed by REFLEX \citep{Bohringer:2001p1461} and NORAS \citep{Bohringer:2000p1456} based on the {\it ROSAT} all-sky survey \citep{Truemper:1993p5035} as well as a by a number of cluster searches in the deep ROSAT archival pointings \citep{Scharf:1997p6157, Rosati:1998p904,Vikhlinin:1998p6138, Romer:2000p1587}. Preliminary cosmological constraints resulted from these studies involving not only cluster counts, but also their 3-D spatial distribution  \citep{Schuecker:2003p1462}; at this stage however, the main observable quantity that was dealt with was the cluster luminosity function, rather than the mass function \citep{Borgani:2001p4880, Mantz:2008p5027}.
Parallel analyses invoking the distribution of temperatures in {\it ROSAT} and {\it ASCA} clusters also provided cosmological constraints, however somewhat debated \citep{Henry:1997p5438, Eke:1998p6187, Viana:1999p6226, Pierpaoli:2001p6231, Pierpaoli:2003p6222,Henry:2004p4668, Henry:2009p4428}, but consistent with findings from the X-ray luminosity function. One of the major shortcomings of these studies rapidly turned out to be lack of reliable mass-observable relations, ideally in the form of scaling relations, and how these would evolve as a function cosmic time.

With the advent of {\it XMM-Newton} and {\it Chandra}, previous samples underwent deep observations  and, in parallel,  the interest in X-ray surveys for cosmological analyses increased. A very significant amount of observing time was devoted to the determination of the cluster scaling relations, for samples a priori thought to be representative of some cluster population (see \citet{Pratt:2009p322} and references therein).
In particular, the 400d survey \citep{Burenin:2007p1251} provided a sample of some 90 clusters selected in the RASS and followed-up by deep {\it Chandra} observations, allowing precise mass measurements based on high-quality X-ray data. The cosmological analysis presented in \citet{Vikhlinin:2009p1250} relies on the mass and redshift distribution of clusters.
\citet{Mantz:2010p1259} used more than 200 X-ray selected clusters, some of them having a deep follow-up, in an analysis that combines the gas mass fraction in clusters with their abundance  per mass and redshift bin. 
Simultaneously, new cluster samples have been assembled either from dedicated XMM surveys \citep{Pierre:2004p6232} or from the XMM and Chandra archival data: e.g.~ChaMP \citep{Barkhouse:2006p3255} and XCS \citep{Romer:2001p6300,Mehrtens:2011p4564}. Corresponding cosmological analyses are still in progress \citep{Sahlen:2009p298}, but one of the major outcome was to realise that selection effects can be as critical as the proper knowledge of the cluster scaling relations. \citet{Pacaud:2007p250} have shown that, unless a high flux limit is assumed, X-ray cluster samples are best characterised by a two-dimensional selection function (analogous to a surface brightness limit). Further, because of the steepness of the cluster mass function and of the (currently poorly determined) dispersion in the scaling laws, these relations appear to be always  biased toward the most luminous objects with respect to the mean, unless a thorough  treatment of the selection function is introduced; this becomes especially challenging as redshift increases.
Ideally, for a given cluster sample, one would need to simultaneously model (i) the selection effects (ii) the scaling relations and (iii) the cosmology.

From the observer's point of view, the bottle neck in building large cosmological samples is the time-consuming optical follow-up to obtain spectroscopic redshifts for each cluster. However, redshift (and mass) information is already encoded in the X-ray spectra of the clusters. In principle, it is possible to make use of this information in a statistical way, even in the low-count regime, thanks to a dedicated formalism \citep{LloydDavies:2010p4565,Yu:2011p3518}.
The goal of the present article is to investigate such a new approach to the cosmological analysis of large samples of X-ray clusters: the CR-HR method.
Conversely to methods requiring redshift information for each cluster and inferring the cluster mass distribution through various X-ray proxies, {\em we handle X-ray instrumental observables only} namely, the count-rates in several energy bands. In the scientific analysis, we self-consistently model the cosmology, scaling laws, selection effects and the instrumental responses to predict count-rate distributions that can be directly compared to the purely observational data.

The structure of this paper is as follows. We begin by presenting the motivations of the CR-HR method and describe its principle. We then give  the key ingredients involved in the  construction of the CR-HR diagram that we illustrate for a shallow XMM survey (Sect.~\ref{ingredients}). Then,  we explain the modeling of measurement errors and their inclusion in the analysis (Sect.~\ref{measurement_errors}). Next, we describe the formalism adopted in our Fisher analysis used to evaluate the CR-HR method; we present the expected constraints for a set of selected parameters (Sect.\ref{fisher}). We discuss and summarize our results in Sect.~\ref{discussion}.  

Throughout the paper, we assume a flat ($\Omega_k=0$) $\Lambda$CDM cosmology with parameters given by WMAP-5 best fit values \citep{Dunkley:2009p3181}.


\section[]{The CR-HR method}
\label{method}

In this work, we consider a shallow X-ray survey and assume that a  robust procedure allows the  construction of well-defined samples of clusters of galaxies. By definition, the survey selection function is based on X-ray observable criteria only. The survey is supposed to be shallow in the sense that a few hundred photons, at most, are collected for each cluster and, thus, may   enable a mean temperature estimate of the intra-cluster medium but no radial temperature profiles.
This is the case for most of the current analyses to date (\citet{Pacaud:2006p3257}, \citet{Sahlen:2009p298}, \citet{Barkhouse:2006p3255}, \citet{Burenin:2007p1251}).
Given a cluster sample, we populate a 2D observable parameter space defined by  the measured  X-ray count-rate (CR) and hardness ratio (HR). CR and HR, which contain (partially degenerate) information on the temperature and the redshift of the clusters, are defined for adequately chosen X-ray bands.
Specifically, we construct a CR-HR diagram  from the selected cluster sample, building a two-dimensional density, which behaves like an X-ray color-magnitude diagram.
In the case where optical (photometric or spectroscopic) redshifts are available for each cluster, we divide the sample in redshift bins and associate a CR-HR diagram to each of these slices leading to a three-dimensional $z$-CR-HR diagram.

Such a diagram can, in turn, be obtained using an {\sl ab initio} formalism: (i) setting a cosmological model, we compute the number of clusters as a function of mass and redshift; (ii) each cluster is ascribed an X-ray temperature and luminosity  as well as a physical characteristic size according to empirical scaling laws; (iii) these quantities are subsequently converted into CR, HR for a given X-ray survey and into an apparent size; (iv) finally, only clusters passing the X-ray selection function are retained, enabling the construction of theoretical CR-HR diagrams.
We compute such ($z$)-CR-HR diagrams for a wide range of cosmological models and possible cluster evolutionary scenarios in order to determine which one is the most likely, by comparing with the observed diagram.

Count-rates are purely instrumental quantities, thus in order to test the ability of the CR-HR method to constrain both the cosmology and cluster evolutionary physics, we need to explicit the calculations for a given X-ray survey instrument such as  {\it XMM-Newton}  or {\it eRosita} \citep{Predehl:2010p5342} for instance.  In the present article, we assume a 100~sq.~deg. {\it XMM} survey performed with the EPIC instruments with a sensitivity of $\sim 10^{-14}$~ergs/s/cm$^2$ in the  [0.5-2] keV band for cluster-type sources, i.e. consisting  of 10 ks exposures. The CR-HR method is evaluated by a Fisher analysis and  its efficiency compared to the traditional method relying on the redshift-mass distribution of clusters. Special care is given to modeling  of measurement errors in the Fisher analysis.

\section[]{Ingredients entering the CR-HR method}
\label{ingredients}

	\subsection{Modeling the CR-HR distribution of clusters}
	\label{model_methode}
\begin{figure*}
	\includegraphics[width=\linewidth]{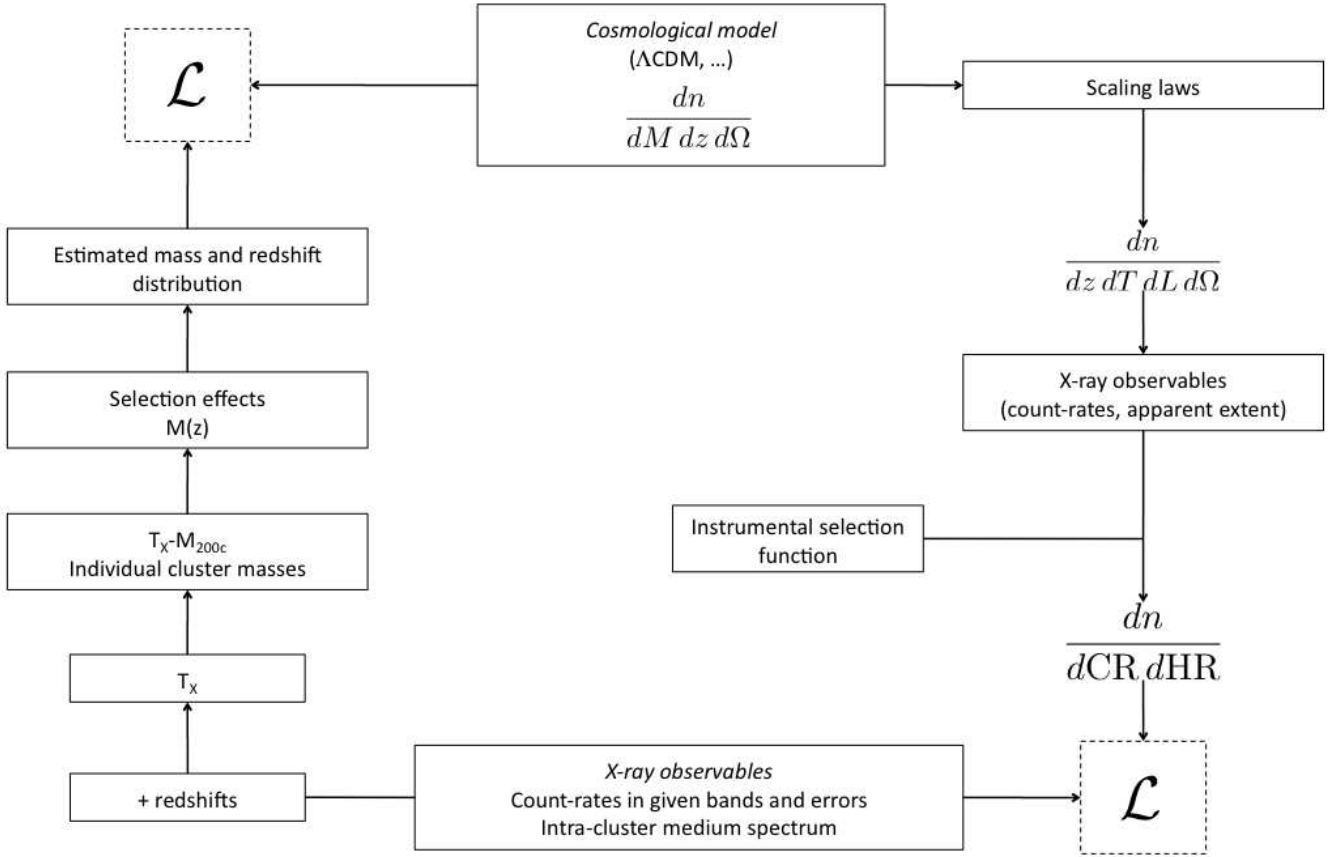}
 \caption{Schematic view of the CR--HR method (right part of the block diagram), illustrating the top-down approach used to link X-ray observables to a cosmological model.
 Left part of the block diagram shows the more traditional method based on individual cluster mass measurements using, e.g. a temperature proxy ($T_X$) and a mass-temperature scaling relation ($T_X-\mdcc$).
 The $\mathcal{L}$ symbol indicates the comparison between model and data (based on, e.g., a minimization of the likelihood function): the CR--HR method compares directly X-ray observables.}
 \label{fig_schema} 
\end{figure*}

This section describes  the steps entering the computation of CR-HR diagrams. We illustrate our method under realistic conditions and qualitatively show its sensitivity to relevant parameters entering the model. Measurement errors  are considered in Sect.~\ref{measurement_errors}. A schematic view of the method is given on Fig.~\ref{fig_schema}.

		\subsubsection{Cosmological mass function}
 We start from a scale-invariant primordial spectrum with slope $n_s=0.961$ and the \citet{Eisenstein:1998p912} transfer function to obtain the $z=0$ power-spectrum $P(k,z=0)$, which is subsequently normalized by $\sigma_8$. The linear power spectrum $P(k,z)$ is evaluated using the redshift-dependent growth factor computed by numerical integration of the partial differential equation. We then compute the rms variance $\sigma(M,z)$ of the field smoothed at a comoving scale $R=(3M/4 \pi \rho_m)^{1/3}$ and inject it into the following functional form describing the differential comoving density of haloes per mass interval $dM$ about $M$ at redshift $z$:
\begin{equation}
\label{equ_massfunction}
\frac{dn}{dM} = f(\sigma) \frac{\rho_m}{M} \frac{d {\rm ln} \sigma^{-1}}{dM}
\end{equation}
where $\rho_m$ is the mean matter density at redshift $z$. We calculate the mass function in terms of $\mdcb$, the mass within a radius $\rdcb$, inside which the mean mass density is 200 times the matter density in the Universe

 We use \citet{Tinker:2008p1554} fit for to obtain $f(\sigma)$ for the corresponding mass definition and then compute the sky-projected, redshift-dependent mass function $dn/d\Omega/d\mdcb/dz$.
The equation of state of dark energy is parametrized through a single parameter $w_0 = P/\rho$, whose value in the case of a cosmological constant is $-1$.

We further transform the mass function in terms of $\mdcc$, defined relatively to the critical density of the Universe. This conversion ($\mdcb$ to $\mdcc$) is motivated by the fact that $\mdcc$ is the mass definition entering our scaling-law formulae (see Section~\ref{ingredients}) and  is performed using the fitting formula from \citet{Hu:2003p153}; for this purpose, we assumed a NFW mass profile \citep{Navarro:1997p3068} and a concentration model from \citet{Bullock:2001p890}.

		\subsubsection{Cluster X-ray emission: scaling laws and brightness profiles}

The X-ray emissivity of clusters basically depends on three quantities: the redshift $z$, the cluster X-ray temperature $T$ integrated over the whole cluster extent, and its total bolometric luminosity $L_X$ (along with some dependence on the metallicity of the ICM).
Scaling relations between cluster masses and these quantities have been extensively studied in the local Universe \citep{Arnaud:1999p398, Arnaud:2005p392, Vikhlinin:2006p393, Pratt:2009p322} down to the low-mass end \citep{Sun:2009p4049}. As usual, we model the cluster scaling relations by power-laws. Given that the  physical processes determining the evolution of these relations are still a matter of debate, we parametrize the evolution by the factor $(1+z)^{\gamma}$ \citep[e.g.][]{Voit:2005p3921}. Our mass-observable relations read:
\begin{equation}
	\label{equ_mt}
   \frac{\mdcc}{10^{14} h^{-1} {\rm M}_{\odot}}  = 10^{\normmt} \Big(\frac{T}{4 {\rm keV}}\Big)^{\powmt} E(z)^{-1} (1+z)^{\zevolmt}
\end{equation}
\begin{equation}
	\label{equ_lt}
   \frac{L_X}{10^{44}\,{\rm ergs/s}}  = 10^{\normlt} \Big(\frac{T}{4 {\rm keV}}\Big)^{\powlt} E(z) (1+z)^{\zevollt}
\end{equation}		
Further, the intrinsic scatter in those relations  is an important ingredient for the modeling the cluster population (e.g. \citealt{Stanek:2006p381},\citealt{Pacaud:2007p250}). We introduce $\scattmt$ and $\scattlt$, the scatter in $T$ at fixed $\mdcc$ (respectively in $L_X$ at fixed $T$) and assume they are independent of redshift, mass and temperature.

Finally,  we assume a surface brightness profile given by a $\beta$-model \citep{Cavaliere:1976p375} with $\beta=2/3$ and a varying core radius $r_c$.
The scaling of $r_c$ with other cluster quantities is complex and depends on the details of the intra-cluster medium physics (see e.g. \citealt{Sanderson:2003p5159}, \citealt{Ota:2004p5088}, \citealt{Alshino:2010p5258}) but it can reasonably assumed that $r_c$ scales with the size of the dark matter halo. We thus take a $\xc = r_c/\rccc$ parameter, constant at all redshift and masses (with $\rccc$ being defined as the radius enclosing a mean density of 500 times the critical density of the Universe).

		\subsubsection{An instrumental model for XMM observations}

\begin{figure}
	\includegraphics[width=84mm]{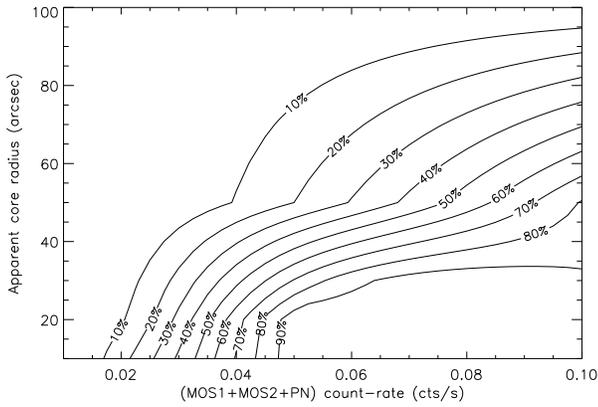}
 \caption{Selection function used throughout this analysis, obtained from realistic simulations of XMM cluster observations  \citep{Pacaud:2006p3257}. The detection probability  is expressed as a function of two observable quantities: the total count-rate collected by the three detectors, and the core radius of the input $\beta$-model ($\beta=2/3$).}
 \label{fig_selfunc} 
\end{figure}

Most of the  cluster detection algorithms in the X-ray waveband are based on a two-step procedure:  source detection is run on a filtered image, followed by   fitting a cluster emission model on the raw photon image, accounting for the Poissonian nature of the signal \citep[e.g.][]{Bohringer:2001p1461, Burenin:2007p1251,Pacaud:2007p250,LloydDavies:2010p4565}. The efficiency of such an algorithm, in terms of completeness and purity, is evaluated by extensive image simulations. This finally enables the determination of cluster selection functions based exclusively on X-ray criteria, which are, in general, more complex than a simple flux limit. Following \citet{Pacaud:2006p3257},   we use a two-dimensional parametrisation involving the count-rate in the [0.5-2]~keV band and the apparent core radius. Figure~\ref{fig_selfunc} shows our adopted selection function (the C1 selection, \citealt{Pacaud:2006p3257}) which corresponds to an uncontaminated cluster sample, for XMM exposure times of the order 10\,ks.

Each cluster is characterized by a redshift $z$, a temperature $T$ and a bolometric luminosity $L_X$. Count-rates are derived from physical fluxes for a given spectral emission model and using the proper  instrumental responses. In this work,  we assume a thermal  plasma model (APEC , \citealt{Smith:2001p6422}) having a metal abundance of 0.3\,$Z_{\odot}$ along with a galactic absorption corresponding to $N_H=3\times10^{20}$~cm$^{-2}$ \citep{Grevesse:1998p5186}. Response matrices for the three EPIC detectors onboard XMM (MOS1, MOS2, \citealt{Turner:2001p5291}; and PN, \citealt{Struder:2001p5293}) and THIN filter are used to produce an observed spectrum  (number of counts collected by second in each energy channel) which is in turn integrated over  specific energy bands to yield the desired countrates.

For the purpose of our analysis, we  define three working energy bands: [0.5-2]~keV (band ``tot", which is also the detection band), [1-2]~keV (band ``1") and [0.5-1]~keV (band ``2"). The choice of these bands is documented in Appendix B.
In all what follows, we assume that we measure  total countrates, i.e. over the full cluster extent (see discussion in App.~\ref{ERRORS})

\begin{figure}
	\includegraphics[width=84mm]{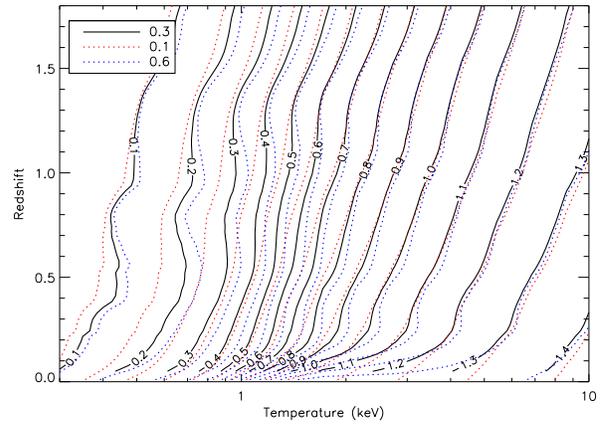}
 \caption{Lines of iso-hardness ratio $\mathrm{CR_1}/\mathrm{CR_2}$ in the plasma temperature-redshift plane for the XMM EPIC instrument. $\mathrm{CR_1}$ is the count-rate in [1-2]\,keV and $\mathrm{CR_2}$ in [0.5-1]\,keV.
An APEC plasma model with abundances 0.3\,$Z_{\odot}$ and a galactic absorption $N_H=3.10^{20}$~cm$^{-2}$ is used. Changes in abundance value from 0.1 to 0.6\,$Z_{\odot}$ are indicated by the dotted coloured lines.}
 \label{fig_example_hr} 
\end{figure}

We define the cluster hardness ratio by $\mathrm{HR}~=\mathrm{CR_1}/\mathrm{CR_2}$ where $\mathrm{CR_1}$ and $\mathrm{CR_2}$ are count-rates measured in [1-2] and [0.5-1]\,keV respectively.
We neglect possible spatial variations of the hardness ratio across the cluster X-ray extent. In practice, for the type of surveys considered here, both the faintness and the small extent of the objects (compared to the instrumental PSF) prevent from resolving such detailed structure.
We thus treat the cluster emission as equivalent to that of a single-temperature plasma and this is consistent with the fact that reference scaling relations have been computed by fitting a single plasma model to various cluster spectra.
For a given spectral model depending on parameters ($z$, $T$, $L_{X}$), the hardness ratio does not depend on luminosity. Figure \ref{fig_example_hr} shows the redshift and temperature dependence of the hardness ratio values :   
a pure bremsstrahlung spectrum would exhibit a degeneracy  of the $T/(1+z)$ type; however the presence of metallic lines entering the energy bands at different redshifts induces more subtle effects, especially at low temperatures where they are prominent.

		\subsubsection{Summary of the model parametrization}
Table \ref{table_parameters} summarizes the main parameters used in our analysis. Beside the WMAP5 cosmological model, parameters governing the cluster M-T and L-T scaling relations are defined by eqs.~\ref{equ_mt} and \ref{equ_lt}. The local $M-T$ relation was taken from \citet{Arnaud:2005p392} using their relation for hot clusters and $\delta=200$.
Following \citet{Alshino:2010p5258} we set $\xc=0.1$, the ratio between the core radius of the X-ray $\beta$-model and $\rccc$.
The local $L_X-T$ relation is taken from \citet{Pratt:2009p322} using their $L_1-T_1$ relation for ``Non Cool Core clusters".
We justify this choice by finding that our fiducial model along with the selection function of Fig.~\ref{fig_selfunc} yields 5.7 clusters per sq.~deg., consistent with the observed density of clusters in the 10\,ks XMM-LSS survey \citep{Pacaud:2007p250}. Choosing their relation for ``All" clusters would lead to a higher density of 10.6 clusters per sq.~deg, indicating an incompatibility between this relation and the XMM--LSS selection function, possibly originating from the different fractions of cool core clusters in the samples under study.
A typical cluster at $z=0.4$ with $\mdcc=10^{14} h^{-1} {\rm M_{\odot}}$ has a temperature $T= 2.2$~keV, a total bolometric luminosity $L_X= 0.55 \times 10^{44}$~ergs/s, and radii $\rccc=0.6$~Mpc and $r_c=11$~arcsec; with a 10 ks XMM exposure, we collect $\sim 500$ photons for this object.

\begin{table*}
	\centering
		\begin{tabular}{@{}cclcc@{}}
\hline
Parameter 	&	Fiducial value	&	Description	&	Ref.	&	Prior	\\
\hline
$\Om$								&		0.249		&&	1	&	No	\\
$\OL$								&		$1-\Om$		&(Flat Universe)&	-	&	-	\\
$\Ob$								&		0.043		&&	1	&	0.003	\\
$\sigma_8$							&		0.787		&&	1	&	No	\\
$w_0$								&		-1			&&	-	&	No	\\
$n_{s}$								&		0.961		&&	1	&	0.014	\\
$h$									&		0.72		&&	1	&	0.026	\\
\hline
$\powmt$							&		1.49		&$M-T$ power-law index				&	2	&	0.17		\\
$\normmt$ 							&		0.46		&$M-T$ logarithmic normalization	&	2	&	0.023		\\
$\zevolmt$							&		0			&$M-T$ evolution index				&	-	&	No		\\
$\scattmt$							&		0.1			&$M-T$ constant logarithmic dispersion		&	2	&	0.064		\\
\hline
$\powlt$							&		2.89		&$L-T$ power-law index				&	3	&	0.21		\\
$\normlt$							&		0.40		&$L-T$ logarithmic normalization	&	3	&	0.026		\\
$\zevollt$							&		0			&$L-T$ evolution index				&	-	&	No		\\
$\scattlt$							&		0.267		&$L-T$ constant logarithmic dispersion		&	3	&	0.058		\\
\hline
$\xc$								&		0.1			&$\beta$-model core radius scaling wrt. $\rccc$	&4	&	No		\\
\hline
		\end{tabular}
		\caption{\label{table_parameters}Fiducial parameters used in this study.
		Last column shows the standard priors used in the Fisher analysis (see sect. \ref{fisher}).
		References: (1): \citealt{Dunkley:2009p3181}, (2): \citealt{Arnaud:2005p392}, (3): \citealt{Pratt:2009p322}, (4): \citealt{Alshino:2010p5258}}
\end{table*}
	
	\subsection{Illustrative examples}
\begin{figure*}
\includegraphics{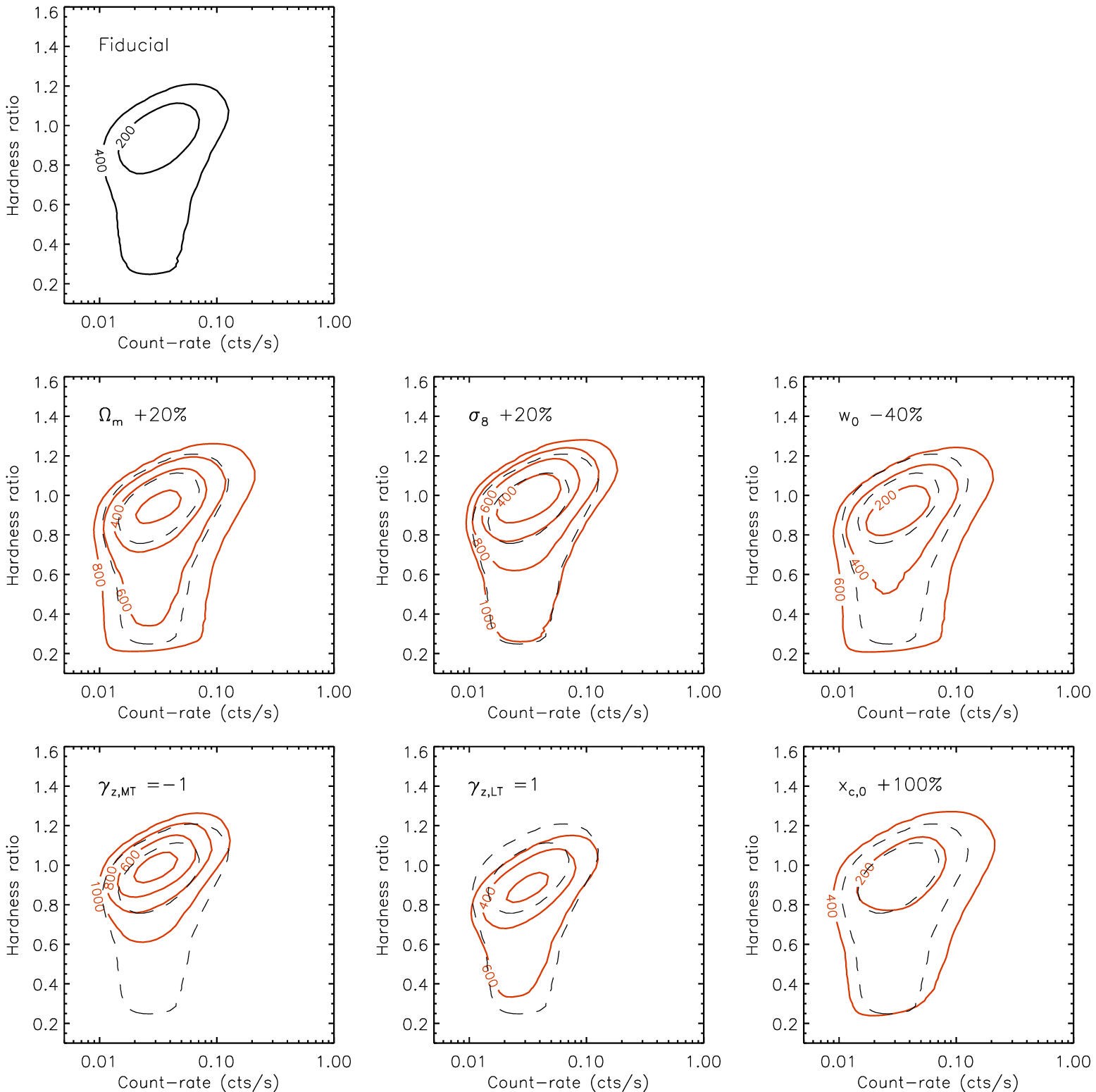}
	\caption{Dependence of the CR-HR diagram on the six free model parameters. The fiducial model (Table~\ref{table_parameters}) is represented by the black-dashed contours and predicts 570 clusters over 100 deg$^2$ (10 ks XMM exposure, C1 cluster selection).
	The red-solid contours represent the model obtained when one parameter at a time is varied.
	The corresponding values for each parameter are: $\Om$~=~0.30 (+20\%), $\sigma_8$~=~0.94 (+20\%), $w_0$~=~-0.6 (-40 \%), $\zevolmt$~=~-1, $\zevollt$~=~1 and $\xc$~=~0.2 (+100 \%).
	Contour levels stand for the number of clusters enclosed by each curve, labelled by steps of 200. The differential quantity $\dndcrdhr$ is constant along each contour.}
	\label{fig_cr_hr_allpar}
\end{figure*}

We show on Fig. \ref{fig_cr_hr_allpar} the  CR-HR distribution computed for our fiducial set of parameters and illustrate the effect of a parameter change on this diagram.
The most obvious effect of modifying one of the parameters is a variation in the total number of observed clusters. This is particularly striking for $\Om$ and $\sigma_8$ which strongly impact the amplitude of the distribution and thus are relatively well constrained by the total number count alone (as pointed out in \citealt{Haiman:2001p4682}, \citealt{Sahlen:2009p298}).
Parameters governing scaling laws ($\zevolmt$, $\zevollt$) enter at the cluster selection stage: increasing e.g. $\zevollt$ increases the luminosity and thus the detectability of clusters at higher redshifts.
Beyond this first-order overall change in amplitude, the shape of the distribution is affected in various ways when the model parameters are varied. If one is able to detect these changes within the measurement uncertainties and systematic errors, degeneracies between parameters can be broken. For instance, a change of 20\% in the value of $\Om$ uniformly changes the amplitude of the CR-HR distribution while a $+1$ modification in $\zevollt$ also shifts the center of the distribution towards lower HR and lower CR.

	\subsection{Generalizing: adding redshift information}
\begin{figure*}
	\includegraphics{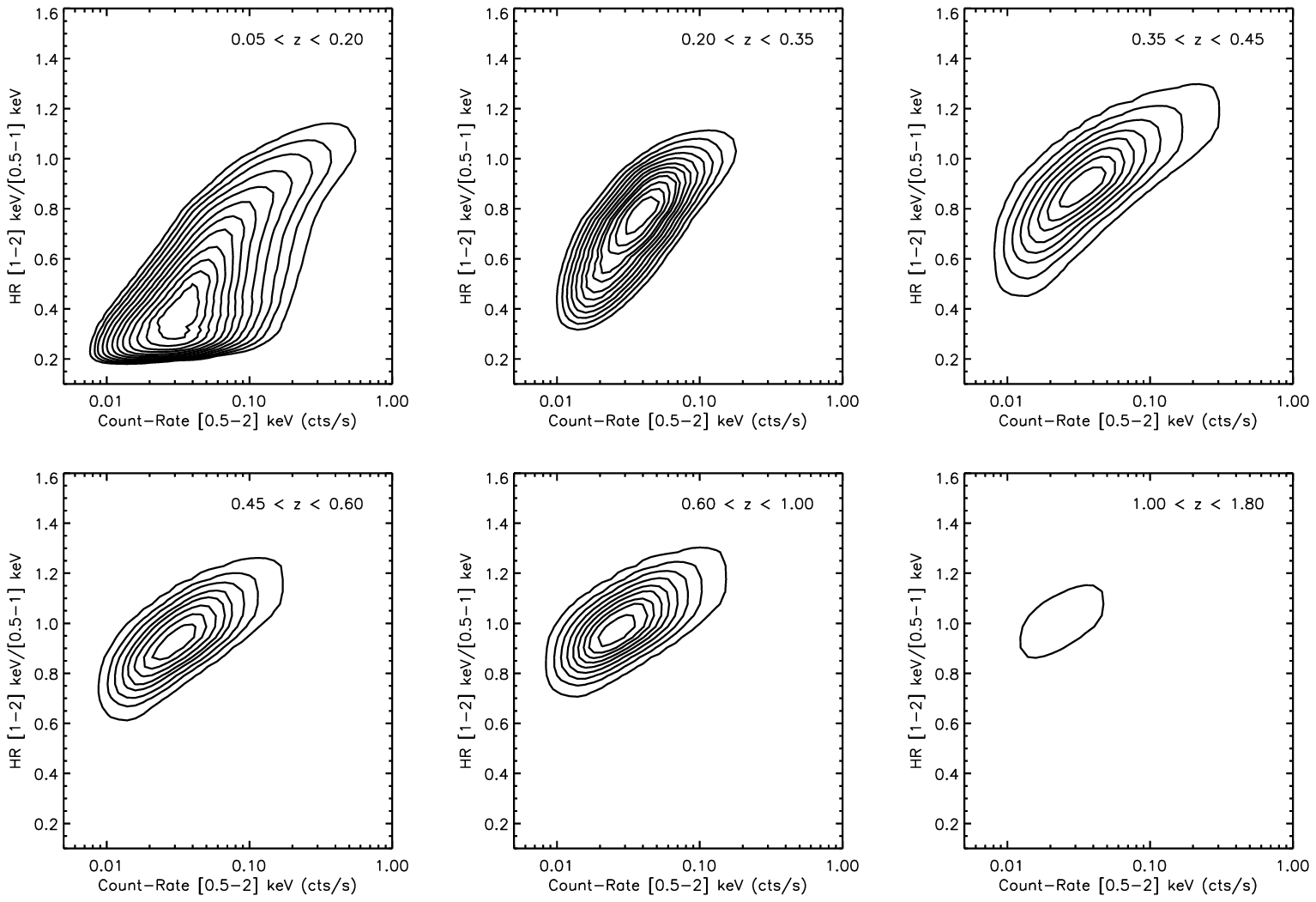}
 \caption{The CR-HR diagram resolved in redshift bins for the fiducial model. Each contour from the innermost to the outermost encloses 10, 20, 30,... clusters detected over 100 deg$^2$ (10 ks XMM exposure, C1 cluster selection). Measurement errors are neglected for this figure. Practically, we compute a 3-dimensional differential density $\dndzdcrdhr$ which is then integrated in defined bins ($\Delta z \Delta {\rm CR} \Delta {\rm HR}$) to provide an histogram of the clusters in the observable space.
 This representation can be built straight from the observable data, without any assumption on cosmological or scaling-laws parameters, and captures the key features of the sample (see text).
 }
 \label{fig_dzdcrdhr} 
\end{figure*}

We also consider the case where the survey benefits from an optical spectroscopic follow-up, providing cluster redshifts. To model this case, we define thin redshift slices and repeat the procedure described above in each of these slices to derive the corresponding tridimensional quantity $\dndzdcrdhr$. The resulting distributions are illustrated  on
Fig. \ref{fig_dzdcrdhr} for given redshift ranges.  Such diagrams almost fully characterize the whole cluster population  using purely observable (instrumental) quantities:  the redshift distribution, the evolution of the CR-HR distribution as a function of redshift and how these quantities are related in the sample. In fact, the $\dndzdcrdhr$ distribution is analogous to the $dn/dz/dL_X/dT$ distribution but can be readily obtained from the available data without any assumption on scaling laws and on cosmological parameters.
	
\section{Accounting for measurement errors}
\label{measurement_errors}

Up to this point we did not include measurement errors arising from the cluster individual measurements. This is however a key issue in the interpretation of the CR-HR diagrams.
In this section, we detail our procedure for modeling the measurement errors in the synthetic distributions. We first describe how count-rate measurement errors impact the CR-HR diagrams. In a second step, we also estimate what would be the uncertainties in  the cluster mass estimates  (based on a M-T proxy)  for exactly the same set of XMM observations.
This step is intended to allow us to eventually  compare the efficiency of the CR-HR method to the traditional method based on cluster masses, in the ideal case where redshifts are available.

	\subsection{Including measurement errors in the CR-HR and $z$-CR-HR diagrams}
CR-HR diagrams involve three measurements for each detected cluster: the wide-band ([0.5-2]~keV) count-rate CR$_{\tot}$, and two narrow-band ([1-2]~keV and [0.5-1]~keV) measurements, CR$_1$ and CR$_2$ such that HR~=~CR$_1$/CR$_2$.
 
Errors on these measurements mostly come from Poisson fluctuations in the signal, from the background level hampering flux measurement up to large projected radii, and from the lack of spherical symmetry amplified by PSF distortion effects. Errors on a measured quantity X knowing the true underlying value $\widehat{X}$ are expressed through a distribution $P(X | \widehat{X})$. For the purpose of this demonstrative paper, we assume a gaussian error model for CR$_{\tot}$, CR$_1$ and CR$_2$, without bias and having a non-constant scatter of the form:

\begin{equation}
\label{equ_error_crhr}
\sigma_{{\rm CR} | \crhat} = \sigma_0 \Big(\frac{T_{\rm exp}}{\rm 10\ ks}\Big)^{-1/2} \Big(\frac{\crhat}{{\crhat}_0\ {\rm cts/s}} \Big)^{1/2}
\end{equation}

\begin{figure}
	\includegraphics[width=84mm]{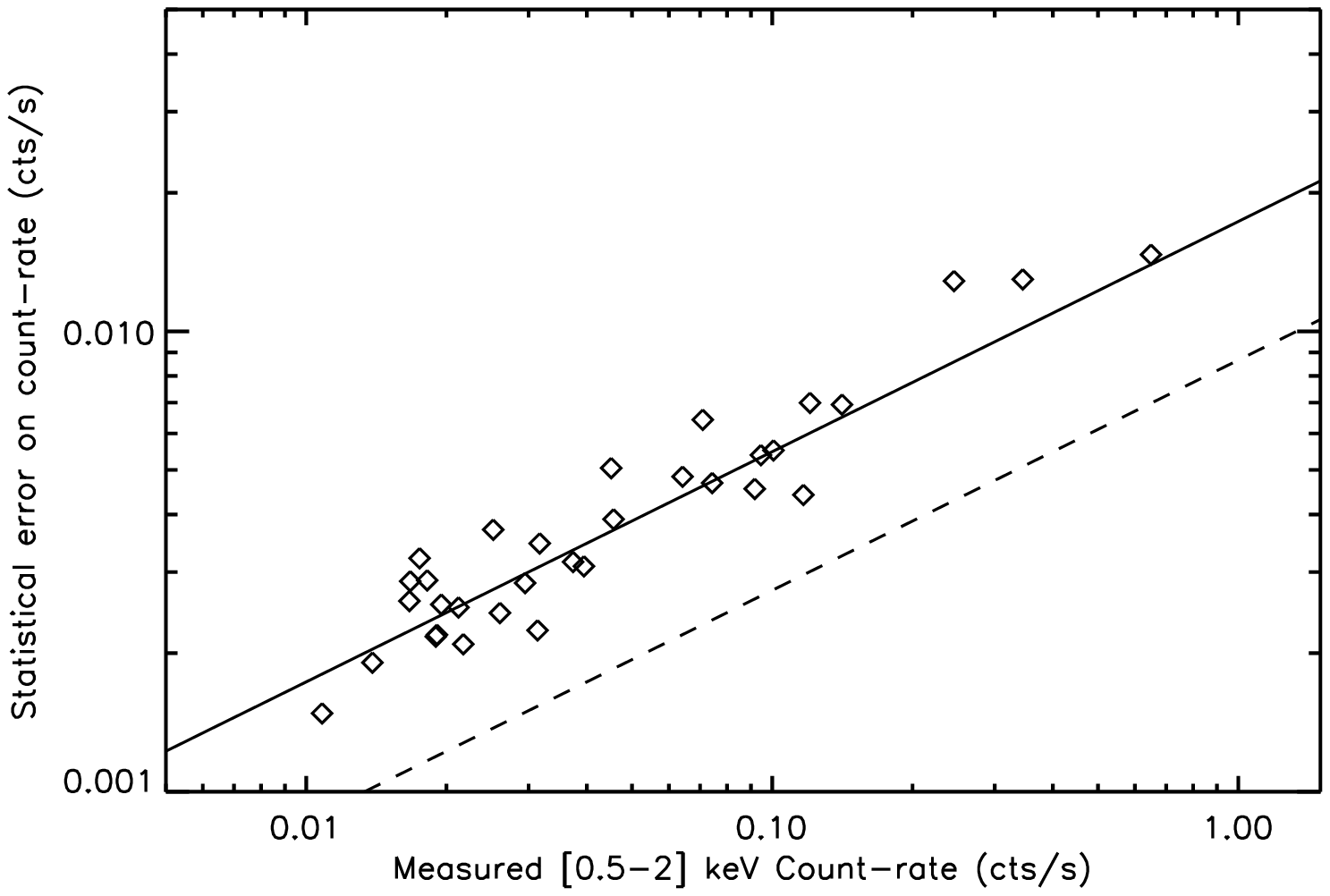}
 \caption{Count-rate measurement errors as a function of the measured count-rate, for the 32 C1 clusters detected in the 10\,ks XMM-LSS field \citep{Pacaud:2007p250}. The plain black line shows the model adopted in Eq.~\ref{equ_error_crhr} for a 10\,ks survey and the dashed line is for 40\,ks exposure time.}
 \label{fig_error_xmmlss} 
\end{figure}

\begin{figure}
	\includegraphics[width=84mm]{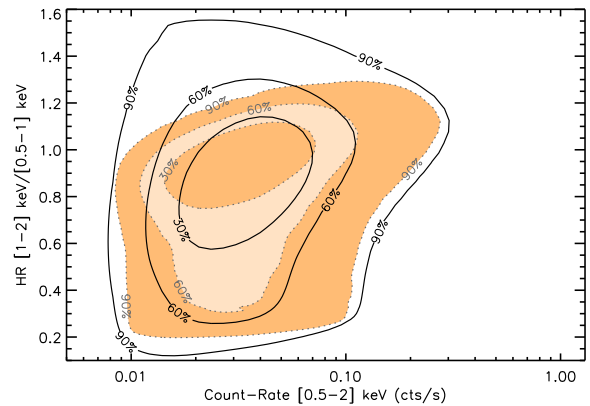}
 \caption{Effect of count-rate measurement errors on the predicted CR-HR distribution.
 	{\it Shaded contours: } Predicted cluster distribution without measurement errors for 10\,ks XMM exposure, C1 cluster selection.
	{\it Unshaded contours: } Same distribution after taking into account measurement errors on both the [0.5-2]\,keV count-rate and the hardness ratio.
	 Contours enclose respectively 30, 60 and 90\% of the total expected number of clusters.
}
 \label{fig_error_crhr} 
\end{figure}

For ${\crhat}_{\tot}$, ${\crhat}_1$ and ${\crhat}_2$ we assume ${\crhat}_0=0.03$~cts/s and $\sigma_0=0.003$\,cts/s (i.e. a 10\% relative error in count-rate measurement for 300 collected photons). This simple model allows to account for the dependence of measurement errors on the number of photons as $\propto \sqrt N$. We checked its validity for measurements of C1 clusters in the 10\,ks deep XMM-LSS field presented in \citet{Pacaud:2007p250}, see Fig.~\ref{fig_error_xmmlss}. We note that including lower flux systems in the diagram would imply a more precise model describing the increased influence of background on these errors.
Errors on ${\rm HR} = {\rm CR}_1 / {\rm CR}_2$ are estimated by simulating $(200)^3$ realistic cluster spectra on a fine (z, $L_X$, T) grid, then computing their (true) ${\crhat}_{\tot,1,2}$ and simulating ${\rm CR}_{\tot,1,2}$ following the gaussian error model presented above. Then, at fixed ($\crhat$,~$\hrhat$) values we compute the standard deviation $\sigma_{{\rm CR,HR} | \hrhat}$ from this set of simulated values.
Figure \ref{fig_error_crhr} shows how $\sigma_{{\rm CR} | \crhat}$ and $\sigma_{{\rm HR}|{\crhat,\hrhat}}$ impact the predicted CR--HR distribution of clusters in the sample. As expected from propagating the errors, the relative uncertainty on HR is larger than that on CR and it increases as the number of collected photons is lower.

Using this error model, we ``blur" the expected $\dndcrdhr$ distribution, in the same way as a varying PSF would affect an image. Practically, this is done by dividing the initial diagram into fine bins, then redistributing the information in each bin into its neighbors using a bi-dimensional gaussian distribution with scatters $\sigma_{ {\rm CR} | \crhat}$ and $\sigma_{{\rm HR}|{\crhat, \hrhat}}$.

In the case where individual cluster redshifts are available, a similar procedure is applied to the three-dimensional $\dndzdcrdhr$ distribution. As count-rates measurements are independent on redshift precision, measurement errors consist of two independent components. Errors on CR and HR are applied on each redshift slice in the same way as for the two-dimensional $\dndcrdhr$ distribution.

Finally, redshift measurement errors are accounted for by narrowing or enlarging redshift bins when integrating the density in cubic cells (see sect.~\ref{fisher}).

	\subsection{Comparison exercise: errors on the estimated cluster masses}
A  traditional method for the cosmological handling of X-ray cluster samples  is to compute a mass proxy for each cluster and, subsequently, analyze the resulting redshift and mass distribution. Since in Sec. \ref{fisher} we will compare the CR-HR method with the traditional approach, we need to model the cluster mass accuracy that is obtainable with exactly the same X-ray information.
Appendix~\ref{massdist} reviews our assumed procedure for deriving  $dn/dM/dz$,  the mass proxy being the cluster X-ray temperature.
Apart from the intrinsic scatter in the M--T relation, errors on the mass determination mainly arise from temperature measurement errors. Considering the parameter set from Table \ref{table_parameters}, we derive at each $(z,\mdcc)$ the expected number of photons $N_{\rm phot}$ collected in the [0.5-10]~keV band with a given exposure time $T_{\rm exp}$. We compute:
\begin{eqnarray}
\label{equ_error_m}
\Delta {\rm ln} \mdcc & \simeq & \powmt \Delta {\rm ln} T \nonumber \\
							& \simeq & \powmt \bigg(\frac{N_{\rm phot}}{400}\bigg)^{-1/2} \frac{\Delta T}{T} \bigg|_{N=400}
\end{eqnarray}
where $\Delta T/T |_{N=400}$ is the relative temperature error that would be obtained with a 400 photons spectrum. This quantity is taken from Fig.~A1 of \citet{Willis:2005p5247}, considering their error bars only.

\begin{figure}
	\includegraphics[width=84mm]{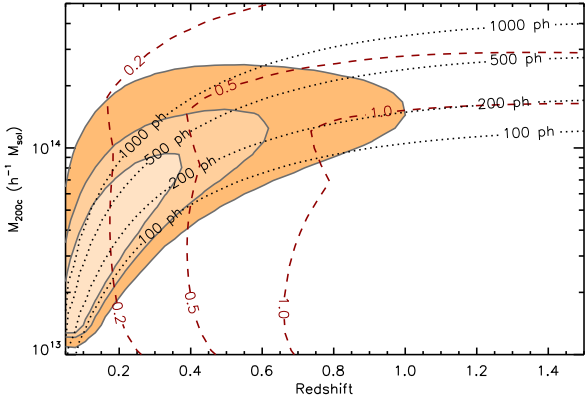}
 \caption{Assumed errors on the measured mass ${\rm ln} \mdcc$ as a function of redshift and $\mdcc$. For each cluster we suppose that $\mdcc$ is obtained by converting the X-ray temperature estimate through the $M-T$ scaling law evolved at the cluster redshift. {\it Red dashed lines: }lines of constant measurement errors on ${\rm ln}(\mdcc)$ in the $\mdcc$-$z$ plane, {\it Black dotted lines: }net number of photons collected in the [0.5-10]\,keV band used for the spectral fit. {\it Shaded contours: }fiducial distribution of detected clusters (10\,ks XMM exposure, C1 selection). Each contour encloses respectively 30, 60 and 90\% of the total expected number of clusters.}
 \label{fig_error_zm} 
\end{figure}

Figure \ref{fig_error_zm} illustrates our projected errors on mass measurements for a 10ks XMM observation. At high redshift, the main source of uncertainty comes from the number of collected photons, and the more massive the cluster, the better the mass measurement. At lower redshifts the relative measurement error is almost independent on the mass for the clusters being studied. This is because the temperature of massive, hot clusters is more difficult to determine as they lack emission features \citep[e.g.][]{Willis:2005p5247}.

Assuming redshifts are poorly determined - i.e. if only photometric redshifts are available -  mass measurements are further degraded. First, for a pure bremsstrahlung spectrum with $T(1+z) \sim $constant at very first order, the spectral fit yields a temperature estimate with a relative dispersion $\Delta T/T \sim \Delta z/(1+z)$.
Second, the conversion from temperature to mass depends on redshift through the $(1+z)^{\zevolmt}$ factor in equation~\ref{equ_mt} and a poor knowledge of $z$ impacts the mass estimate.
We neglect the latter source of uncertainity as our fiducial model is computed at $\zevolmt = 0$. The former is added in quadrature to the statistical error described in eq.~\ref{equ_error_m}), and we take $\Delta z/(1+z)=0.07$ when considering photometric redshifts.

\section[]{The Fisher analysis}
\label{fisher}

In this section, we evaluate the level of performance of the method based on the knowledge of  $\dndcrdhr$ and $\dndzdcrdhr$. We quantify this performance in terms of constraints on cosmological parameters and on scaling laws related parameters.
We describe the Fisher formalism used in this analysis and its results. In the ideal case where redshifts are available for each cluster, we compare the efficiency of the CR-HR method with the traditional approaches using $dn/dz$ and $dn/dM/dz$.

	\subsection{Fisher formalism}
The principle of Fisher matrices applied to cosmological forecasts is thoroughly discussed in \citep[e.g.][]{Tegmark:1997p2228,Eisenstein:1999p5256,Heavens:2009p5254}. Here we briefly recall the approach and show how we applied it to evaluate our method.

Given a set of measured observables $\{D_1,...,D_n\}$ assumed to be uncorrelated, a parametric analysis aims at constraining a set of parameters $\{\theta_1,...,\theta_p\}$ under a physical model $\mathcal{M}$. Defining the likelihood $\mathcal{L}=P(D_i | \theta_{\mu},\mathcal{M})$ and assuming a prior distribution $P(\theta_{\mu}|\mathcal{M})$, the posterior $P(\theta_{\mu}|D_i,\mathcal{M}) \propto \mathcal{L} \times P(\theta_{\mu}|\mathcal{M})$ contains all the information needed to derive confidence intervals on the $\theta_{\mu}$.
If we denote by $\obs_i(\theta_{\mu})$ the observable predicted by the model and assuming Poisson distribution in each bin $i$, the likelihood reads:
\begin{eqnarray}
\label{equ_likelihood}
{\rm ln}\mathcal{L} & = &\sum_i {\rm ln} P_{\rm Poiss} (D_i | \obs_i(\theta_{\mu})) \\
					& = &\sum_i \big(-\obs_i(\theta_{\mu}) + D_i {\rm ln}\obs_i(\theta_{\mu}) - {\rm ln} D_i ! \big)
\end{eqnarray}

Defining the Fisher matrix as:
\begin{equation}
F_{\mu \nu} \equiv - \Big\langle \frac{\partial^2 {\rm ln} \mathcal{L}}{\partial \theta_{\mu} \partial \theta_{\nu}}\Big\rangle ,
\end{equation}
one obtains under those assumptions:
\begin{equation}
\label{equ_fisher}
F_{\mu \nu} = \sum_i \frac{1}{\obs_i} \frac{\partial \obs_i}{\partial \theta_{\mu}} \frac{\partial \obs_i}{\partial \theta_{\nu}}
\end{equation}

Marginalized parameter uncertainties as well as their mutual correlations are encoded in the covariance matrix $C_{\mu \nu} = F_{\mu \nu}^{-1}$. For instance, the 1-$\sigma$ marginalized error on parameter $\theta_{\mu}$ is given by $\sqrt{C_{\mu \mu}}$.
External gaussian priors on parameters can be included by simply adding together Fisher matrices. Particularly, if $\theta_1$ has a prior $\sigma_1$, the resulting Fisher matrix is obtained by adding $1/\sigma_1^2$ to the $F_{11}$ term of the original matrix.

We insist on the fact that Fisher matrices only provide the best constraints attainable by the experiment and neglect all terms above linear order. The derived constraints must thus be seen as indicative for e.g. a comparison of two distinct methods. Moreover, a Fisher analysis is valid around a given model and all constraints derived from the matrix inversion depend on the assumed model. In our case, all methods are compared using the fiducial model presented in Table \ref{table_parameters}.
Finally, we note that the derivation of eq.~\ref{equ_fisher} presented above is valid only if $D_i$ does not depend  on $\{\theta_{\mu}\}$. In some cases the computation of the data set $D_i$ requires the knowledge of some parameters among $\{\theta_{\mu}\}$ (e.g. $\Om$ is needed to compute cosmological distances entering the conversion from flux to luminosity). This problem can be partially overcome by predefining a set $\{\theta_{\rm ref}\}$ (it can be the same as $\{\theta_{\rm fiducial}\}$), deriving $D_i$ with this reference set and then correcting $\obs_i(\theta_{\mu})$ so as to compare both values in the same reference space. This is typically the case when deriving constraints from the mass distribution of clusters, since the mass derivation relies on several key parameters of the analysis (see App.~\ref{massdist}).

\begin{figure}
	\includegraphics[width=84mm]{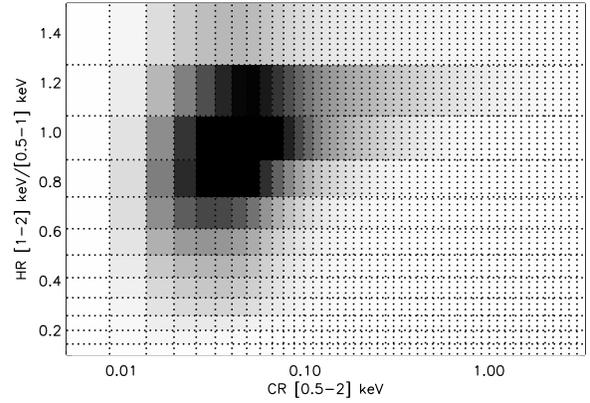}
 \caption{Predicted integrated density $\dndcrdhr$ illustrating the binning scheme applied in our Fisher analysis (see sect. \ref{fisher}). The original distribution is first ``blurred" according to the measurement errors and then binned into cells. Each bin is approximately as large as the measurement error. The model parameters are those from Table \ref{table_parameters}. Shading from white to black represent the expected number of clusters in each cell (white = 0, black = 10 or more clusters). The total number of clusters is 570 over 100 deg$^2$. The error model is defined by equation \ref{equ_error_crhr} with $T_{\rm exp}=10$~ks}
 \label{fig_crhr_integ} 
\end{figure}

Our predicted observable is built from one of the predicted densities $\dndcrdhr$, $\dndzdcrdhr$, $dn/dz$ and $\dndzdm$. Measurement errors are applied following the procedure described in Sect.~\ref{measurement_errors}. A binning scheme is then defined and held fixed, and the $\obs_i$ are defined as the cell-integrated densities.
Binning grids are chosen so that the bin size at each point is approximately as large as the 1-$\sigma$ error size at the considered point. In such a way, correlations between bins are minimized (we do not consider the effect of sample or cosmic variance here, see App.~\ref{ERRORS} for a discussion).
We consider two redshift binning :  $\Delta z=0.1$ and $\Delta z=0.03$.
Observable ranges should span the entire cluster population and we choose:
$z \in [0.05, 1.8]$,
${\rm CR} \in [0.005, 3.5]$ cts/s, 
${\rm HR} \in [0.1, 1.55]$ and 
$\mdcc \in [10^{13}, 3.10^{15}]$ $h^{-1} {\rm M}_{\odot}$.
Figure \ref{fig_crhr_integ} shows a typical example of a CR-HR integrated density. Comparing it to Figure~\ref{fig_cr_hr_allpar}, a substantial amount of information has been lost by including measurement errors and binning the distribution, but the main characteristics of the distribution are still present, in particular its normalization which is the total expected number of clusters in the sample.

Derivatives of the predicted observables with respect to model parameters are evaluated using the five-point stencil approximation:
\begin{eqnarray}
\frac{\partial \obs}{\partial \theta_{\mu}} \simeq
\frac{2}{3} \frac{\obs(\hat \theta_{\mu}+\delta\theta_{\mu}) - \obs(\hat \theta_{\mu}-\delta\theta_{\mu})}{\delta \theta_{\mu}} \nonumber \\
+\frac{\obs(\hat \theta_{\mu}-2 \delta\theta_{\mu}) - \obs(\hat \theta_{\mu}+2\delta\theta_{\mu})}{12 \delta \theta_{\mu}}
\end{eqnarray}
with steps 5\% of the fiducial value for non-zero fiducial parameters and 0.05 for the other parameters ($\zevolmt$ and $\zevollt$).

	\subsection{Results}

In the following, one assumes a 100 sq.~deg. survey uniformly covered by a 10~ks XMM integration, thus leading to a selection function given in Fig.~\ref{fig_selfunc}.
The fiducial model is $\Lambda$CDM with parameters given in Table~\ref{table_parameters}. Thus we always consider a sample of 570 clusters (see Sect.~\ref{ingredients}).

The analysis involves 15 varying parameters: $\{\Om,\Ob,\sigma_8,w_0,n_s,h\}$, $\{\powmt,\normmt,\zevolmt,\scattmt\}$, $\{\powlt,\normlt,\zevollt,\scattlt\}$ and $\xc$. Gaussian priors are applied following Table~\ref{table_parameters}. These priors are uncorrelated, unless for $\Ob$, $n_s$ and $h$ for which correlations from WMAP-5 are taken into account \footnote{As computed from the the Monte Carlo Markov Chains available at http://lambda.gsfc.nasa.gov/}.
We highlight constraints obtained on $\{\Om,\sigma_8,w_0,\zevolmt,\zevollt,\xc \}$ after marginalization over the 9 remaining parameters.

		\subsubsection{Effect of measurement errors}
		\label{errz}
\begin{figure}
	\includegraphics[width=84mm]{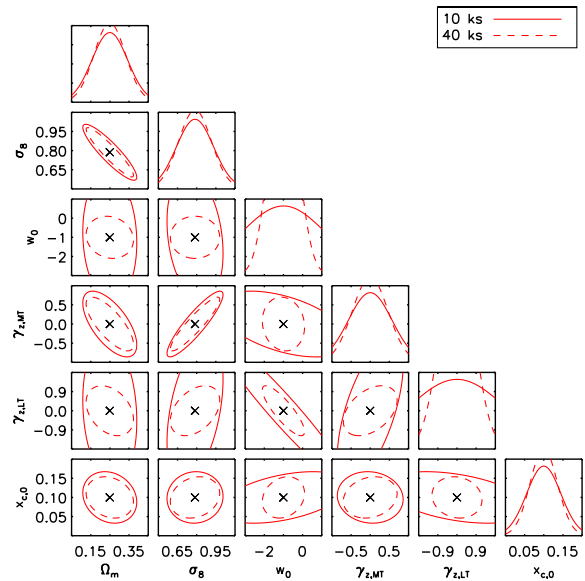}
 \caption{Effect of the count-rate precision on the constraints obtained by the CR-HR method. Displayed are the constraints on the six unknown parameters from Table~\ref{table_parameters}, as obtained by the Fisher analysis. Each ellipse encloses the 68\% confidence area of the marginalized posterior distribution. This figure shows how the uncertainty on each parameter can be tightened by reducing measurement errors by a factor of $\sim 2$ (i.e. with a 40~ks XMM follow-up of each cluster). On the diagonal is shown the marginalized gaussian posterior distribution normalized so as to yield a total probability equals to 1.}
 \label{fig_crhr_depth_dependence} 
\end{figure}

In a first step, we study how the precision on count-rate measurements impacts the constraints on model parameters.  We consider two situations : i) Count-rates are measured on the 10~ks survey data and ii) Improved accuracy is provided by a subsequent 40~ks X-ray follow-up on each detected cluster. (for both cases we have identical selection functions, i.e. the same cluster  sample).
Figure \ref{fig_crhr_depth_dependence} shows the results for these two cases when redshifts are not available, thus using CR-HR diagrams only. 
Constraints on $\Om$, $\sigma_8$, $\zevolmt$ and $\xc$ show little improvement when dividing measurement errors by a factor of two (i.e. going from 10~ks to 40~ks observations).
On the other hand, constraints on $w_0$ and $\zevollt$ are divided by a factor $\sim 3$. This is a consequence of the deformation they imprint on the $\dndcrdhr$ surface, which is better captured in the presence of precise measurements.

\begin{figure}
	\includegraphics[width=84mm]{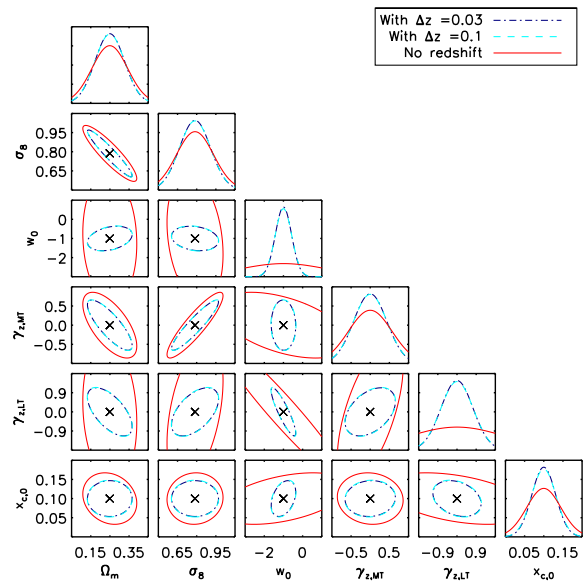}
 \caption{Effect of redshift precision on the constraints obtained by the ($z$-)CR-HR method. Two redshift binnings (accuracies) are considered:   $\Delta z=0.1$ and $\Delta z=0.03$. Red ellipses are the same as on fig.~\ref{fig_crhr_depth_dependence} and show how those constraints are affected by removal of the redshift information.}
 \label{fig_crhr_redshift_knowledge} 
\end{figure}

Further, in the case where cluster redshifts are available, we investigate the impact of the redshift precision on the $z$-CR-HR method.  For this purpose we consider a 10~ks survey i) without redshifts, ii) with very approximate redshifts  and iii) with photometric-like redshift accuracy. Practically, these configurations are rendered by narrowing the redshift bins from $\Delta z=\infty, 0.1, 0.03$ when computing the Fisher matrix. Results are displayed in Fig. ~\ref{fig_crhr_redshift_knowledge}. As expected, we notice an overall improvement of the constraints obtained on model parameters with increasing redshift accuracy. Adding redshift information substantially improves the precision on $w_0$ and $\zevollt$, by a factor of 5 (resp.~3); the other parameters also show an improvement.
However,  refining the redshift bins does not have a strong impact on the results, and a $\Delta z = 0.1$ binning contains almost the full constraining power of the method.

Table~\ref{table_effect_meserror} summarizes the 1-$\sigma$ marginalized uncertainties on the six parameters in the configurations presented above.

		\subsubsection{Comparison with dn/dz}
For comparison purpose, we also quote in Table~\ref{table_effect_meserror} constraints obtained from an analysis that would only involve the redshift distribution of clusters in the sample, i.e. the standard $dn/dz$ cluster counts not making use of the spectral information potentially available in the X-ray data:
even if those properties have an implicit impact on the observed $dn/dz$ (through the survey selection function), they do not interfere with the construction of the redshift histogram of the sample.
Thus, one expects degeneracies between parameters to be important and marginalized uncertainties on individual parameters to be large, if not meaningless. We discuss in App.~\ref{compar_xxl} the comparison between our implementation of the $dn/dz$ analysis and a slightly different method based on the simultaneous fit of the redshift histogram and the L-M relation in several redshift bins as presented in \citet{Pierre:2011p5484}.

\begin{figure}
	\includegraphics[width=84mm]{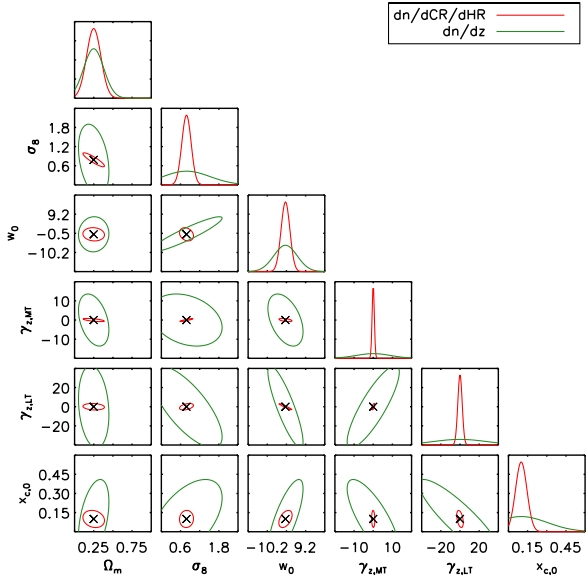}
 \caption{Comparison between constraints obtained from a traditional $dn/dz$ analysis and our proposed observable $dn/d{\rm CR}/d{\rm HR}$, the latter not involving direct redshift measurements, contrary  to $dn/dz$. The redshift bin size for $dn/dz$ is $\Delta z=0.03$ and measurement errors are computed for a 10 ks survey  for $dn/d{\rm CR}/d{\rm HR}$ (see fig.~\ref{fig_error_crhr}). Plotting ranges have been widened to ease visualization, as in most of the cases the contour corresponding to $\dndcrdhr$ is very small in comparison to the corresponding green ellipse.}
 \label{fig_zcrhr_vs_all} 
\end{figure}

Figure~\ref{fig_zcrhr_vs_all} illustrates the comparison between this method (based on $dn/dz$) and our method ($\dndcrdhr$) which does not make use of individual cluster redshifts.
It turns out that for $dn/dz$, parameters $w_0$, $\zevolmt$ and $\zevollt$ are totally unconstrained and $\Om$, $\sigma_8$ and $\xc$ ellipses are considerably widened.
Moreover, additional degeneracies between parameters arise and also participate in diluting the constraints. This is particularly true for both parameters $\zevolmt$ and $\zevollt$ whose effects cannot be disentangled by the redshift distribution only.
As expected the best strategy is thus to use all information available in the survey (redshifts and X-ray measurements) as they help in breaking degeneracies related to the selection function.

\begin{table}
	\centering
		\begin{tabular}{@{}lcccccc@{}}
\hline
Obs.	&	\multicolumn{2}{c}{$\dndcrdhr$}		&	\multicolumn{2}{c}{$\dndzdcrdhr$}		&	\multicolumn{2}{c}{$dn/dz$}	\\
\hline
Depth	&	10~ks			&	40~ks		&	\multicolumn{2}{c}{10~ks}				&	\multicolumn{2}{c}{-}	\\
\hline
$\Delta z$	&	\multicolumn{2}{c}{-}		&	0.1		&	0.03						&	0.1				&	0.03	\\
\hline
Param.			\\
$\Om$		&	0.09			&	0.08			&		0.08		&		0.08			&		0.15		&		0.13		\\
$\sigma_8$	&	0.14			&	0.13			&		0.12		&		0.12			&		0.98		&		0.74		\\
$w_0$		&	2.2				&	0.73			&		0.43		&		0.42			&		9.0(*)		&		5.9(*)		\\
$\zevolmt$	&	0.57			&	0.46			&		0.44		&		0.44			&		11(*)		&		9.0(*)		\\
$\zevollt$	&	2.3				&	0.78			&		0.76		&		0.75			&		41(*)		&		28(*)	\\
$\xc$		&	0.04			&	0.04			&		0.03		&		0.03			&		0.33		&		0.18		\\
\hline
		\end{tabular}
		\caption{\label{table_effect_meserror}
			 Marginalized 1-$\sigma$ constraints on the six cosmological and scaling laws-related parameters for an XMM 100 sq.~deg. survey at 10~ks depth providing a sample of 570 clusters.
			 We show results for three different observables and different measurement errors on CR (count-rate), HR (hardness ratio) and $z$ (redshift).
			 The 40~ks indication on the second row refers to the depth of a potential X-ray follow-up on individual clusters (10~ks meaning no additional follow-up). The $dn/dz$ analysis is independent on any X-ray follow-up..
			The increase in precision of redshift measurements is rendered by narrowing the redshift bins $\Delta z$.
			(*)these parameters can be considered as completely unconstrained by the observable in this configuration.
			}
\end{table}

	\subsubsection{Comparison with $\dndzdm$}
\begin{figure}
	\includegraphics[width=84mm]{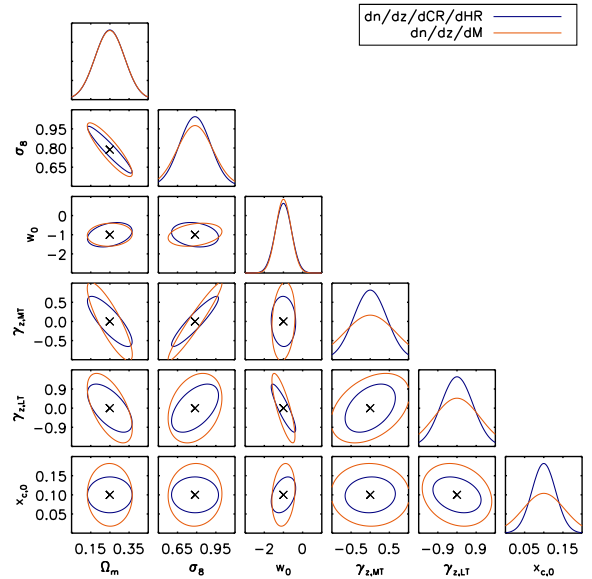}
 \caption{Comparison between constraints obtained from our proposed $z$-CR-HR method (blue ellipses) and from a $dn/dz/d\mdcc$ analysis (orange ellipses) as from Table \ref{table_dndzdcrdhr_dndzdm}.  The mass $\mdcc$ is estimated from the X-ray temperature $T$ then converted via an assumed $M-T$ scaling relation (see Appendix.~\ref{massdist} for details.) The redshift binning is such that $\Delta z=0.03$.}
 \label{fig_dndzdm_vs_dzdcrdhr_spectro} 
\end{figure}

We now assume that redshifts are available for each cluster, at a sufficient precision to allow a binning size of $0.03$, corresponding roughly to photometric redshift precision.
We compare two ways of analysing the data: either directly using the observed quantities ($\dndzdcrdhr$) or using a mass proxy ($\dndzdm$) as described in App.~\ref{massdist}.
Neglecting measurement errors, intrinsic scatter in the scaling relations and systematics, we expect constraints on model parameters to be of the same order of magnitude, as both methods use identical datasets and rely on the same underlying quantity (the ``cosmological" mass and redshift distribution of haloes).
Figure~\ref{fig_dndzdm_vs_dzdcrdhr_spectro} compares the efficiency of the two methods, taking into account measurement errors as presented in Figures~\ref{fig_error_crhr} and~\ref{fig_error_zm}.
Corresponding marginalized constraints are presented in Table~\ref{table_dndzdcrdhr_dndzdm}. An additional column gives the results that would be obtained with a mass precision  10\%, hence the ultimate constraints attainable with the traditional $\dndzdm$ function.

Interestingly, the accuracy reached on parameters of purely cosmological origin ($\Om$, $\sigma_8$ and the dark energy parameter $w_0$) is comparable for the two methods. The relative precision on $\Om$ is about 30\%, while $\sigma_8$ is constrained to $\sim$~15\% in all considered cases, showing slight improvements when reducing measurement errors and narrowing the redshift binning.
On the other hand, a substantial gain is obtained on both parameters $\zevolmt$ and $\zevollt$ governing the evolution of scaling laws with redshift. Using the same data set, a factor $\sim$~1.5 appears when using the $z$-CR-HR representation instead of the $z-\mdcc$ distribution.
Finally, the parameter $\xc$ governing the scaling of the $\beta$-model core radius $r_c$ with $\rccc$ is  constrained twice as well with $\dndzdcrdhr$, up to a relative precision of $\sim$~30\%.

As already mentioned in Sec. \ref{errz}, switching from a redshift accuracy of 0.03 to 0.1   has not  strong impact on the final derived constraints for both methods.

\begin{table}
	\centering
		\begin{tabular}{@{}lccccc@{}}
\hline
Observable	&	(A)						&		(A)					&		(B)					&		(B)				&	(B)		\\
\hline
Depth		&	10~ks					&		10~ks				&		10~ks				&		10~ks			&	$\Delta M/M~=~10\%$		\\
\hline
$\Delta z$	&	0.1						&		0.03				&		0.1					&		0.03			&	0.03			\\
\hline
Parameter			\\
$\Om$		&	0.08					&		0.08				&		0.08				&		0.08			&		0.07	\\
$\sigma_8$	&	0.12					&		0.12				&		0.14				&		0.14			&		0.12	\\
$w_0$		&	0.43					&		0.42				&		0.41				&		0.40			&		0.23	\\
$\zevolmt$	&	0.44					&		0.44				&		0.69				&		0.68			&		0.59	\\
$\zevollt$	&	0.76					&		0.75				&		1.1					&		1.1				&		0.67	\\
$\xc$		&	0.03					&		0.03				&		0.06				&		0.05			&		0.05	\\
\hline
		\end{tabular}
		\caption{\label{table_dndzdcrdhr_dndzdm}
			 Marginalized 1-$\sigma$ constraints on the six cosmological and scaling laws parameters for an XMM 100 sq.~deg. survey at 10~ks depth providing a sample of 570 clusters.
			 We show results for two different observables: (A) is based on the three-dimensional $\dndzdcrdhr$ diagrams and (B), the traditional method, relying on the two-dimensional mass and redshift distribution.
			 The survey depth is 10~ks for both the detection and the measurements. In the last column an uniform mass precision of 10\% is assumed.
			The narrowing of the redshift bins $\Delta z$ renders the increase in precision of redshift measurements.
			}
\end{table}

\section[]{Summary and conclusions}
\label{discussion}

This paper discusses the efficiency of a new method based on strictly observable quantities (i.e. instrument dependent measurements) to analyse the cosmological content of large X-ray cluster samples. Specifically, for each cluster we only make use of an X-ray count-rate, CR, and  of a hardness ratio, HR, plus the cluster redshift, if available ; the 2D (3D) distribution of the CR-HR-(z)  values from the cluster sample constitutes the quantity to be analysed. Compared to the traditional approaches based on the $dn/dz$ and $dn/dz/dM$ statistics, our method follows a purely top-down  procedure, requiring to derive from an {\sl ab initio model} the expected CR-HR diagram. The method constrains in a self-consistent manner the three main ingredients of the model: (1) the cosmology, (2) cluster scaling laws and their evolution and (3) the selection effects inherent to the survey design. Moreover, it avoids the tedious intermediate steps involved in the derivation of the cluster mass estimates for the traditional methods (for instance : spectral fitting, determination of the mean cluster temperature and finally, mass estimate via a scaling relation).
The present study has been done for the particular case of a 100 deg2 survey uniformly paved with 10 ks XMM  observations. We discuss below our results and the assumptions made throughout this work.

\subsection{Main results}
- The CR-HR method is primary intended for the analysis of X-ray cluster surveys for which no information is available on individual cluster redshifts (either from optical spectroscopy or the X-ray observations are too shallow to yield an X-ray spectroscopic redshift); this is the case in the early phase of surveys covering a large fraction of the sky  (e.g. pointed observations from  archival data).
In such a case, the usual approach is to consider the logN-logS distribution of the cluster fluxes, in one or, possibly, in two or more bands. However the constraints one ought to put on the cosmology and cluster physics are at this stage, rather degenerate. By studying the CR-HR diagrams, we make use of all X-ray available information (as far as allowed by the statistical significance of the HR), that is: not only we are able to say `` We have so many clusters in this flux range and in these detection bands'', but also ascribe an X-ray colour and countrate to each cluster, which is much more constraining. At the same time, our top-down procedure avoids the non universal {\it count-rate}\,$\Rightarrow$\,{\it flux} translation step, which is mandatory for studies based on the logN-logS.\\
- The power of the method can be qualitatively intuited from Fig.~\ref{fig_example_hr} as follows: The traditional way of reading this figure leads to the trivial conclusion that redshift is degenerate with temperature, when only an X-ray hardness ratio is available: lines of iso-hardness ratio are almost vertical. Conversely, one can use this property to infer that the HR gives a rough indication of the temperature, independently of the redshift; in parallel, for a given temperature the CR in the  [0.5-2] keV  band provides the normalisation of the spectrum, which depends on the total cluster emissivity (i.e. luminosity) and on the distance of the cluster. Assuming a standard M-L relation, it is thus possible to roughly infer estimates for the mass and the redshift of a cluster knowing the CR and the HR.   \\
- Practically, our method requires to compute a grid of CR-HR diagrams to explore all the parameter ranges one is aiming to constrain (cosmology, cluster physics and evolution) and to find which is the one that best fits the observed diagram. 
To evaluate the actual power of the method, we performed a Fisher analysis which first required  a realistic modeling of how measurements errors on CR and HR dilute the information contained in the diagrams. 
In parallel, for comparison, we performed a similar analysis for a study which would be based on exactly the same X-ray data, but would determine and use the traditional $dn/dz$ and $dn/dz/dM$ distributions. In all of this work, because only a few hundreds of clusters are available, we assume that the local M-T and L-T scaling relations are known ; but this condition can be easily relaxed in the case of all-sky surveys (see paper II). We parametrize the evolution of the scaling laws by two factors $(1+z)^{\gamma}$.\\
- The  calculations presented in this article have been performed in the case
of an XMM/EPIC survey. They can be easily extended to any X-ray telescope
providing comparable spectral-imaging capabilities, i.e. a spectral
resolution of the order of  5-10\% between 0.5-2  kev.\\
- We summarize our results as follows:
\begin{itemize}
\item The CR-HR method (not requiring redshifts) allows a competitive and almost readily available analysis of X-ray surveys without the need to wait for spectroscopic follow-up of collected clusters. It appears much more efficient than the
dn/dz statistics (requiring redshift) and thus, is a very significant improvement over the logN-logS approach.
\item Refining the precision on CR and HR by multiplying the X-ray depth by a factor of 4 (without changing the total number of clusters in the sample) does not significantly impact the determination of $\sigma_{8}$ and $\Omega_{m}$, which are mostly dependent on cluster counts. But a significant improvement is observed for the cluster evolution and, interestingly, on the $w$ parameter of the dark energy equation of state.
\item We further investigated the CR-HR method, by assuming that redshifts are available for all clusters. This allows us to add a third dimension to the diagram:  for a redshift accuracy of $\Delta z \sim 0.1$, we observe a significant improvement  especially for the cluster evolution  and the dark energy parameters. Increasing the accuracy to  $dz \sim 0.03$ does not result in a further improvement, because this is below the cluster evolution time-scale.
\item We finally compare the CR-HR-z method to the dn/dz/dM statistics. Both approaches appear relatively equivalent for the cosmological parameters, while again the CR-HR-z method better constrains the cluster evolution parameters.
\item In all situations, the CR-HR(-z) approach appears to be uniquely well suited to constrain the cluster characteristic size.
\end{itemize}

In conclusion, the CR-HR method appears optimally suited to the analysis of the X-ray data in a survey, and the CR-HR-$z$ method is significantly more efficient than the standard approaches based on the  $dn/dz$ and $dn/dz/dM$ statistics and simpler to implement. We have attempted to make an account as realistic as possible of the various sources of uncertainty entering the method. This is however, probably not exhaustive and we discuss further in Appendix \ref{ERRORS} a number of pending issues in this respect.

	\subsection{Future work}
In a future work we will study the impact of the shape of the selection function on the efficiency of our method. We will also quantify the effect of the scatter in the scaling relations, especially focusing on the conversion from observable to mass in the $\dndzdm$ analysis.
Another point of interest will be to consider how flux measurements in fixed apertures can help in a better determination of model parameters, and in the case of a much deeper survey, how useful it would be to introduce a second hardness ratio pertaining to the harder part of the cluster spectrum. In the latter case we shall also consider introducing $Y_X$ as a proxy for the observable to mass conversion \citep{Kravtsov:2006p387}.

A major, practical, advantage of our method is that there is no need to derive individual masses of detected clusters and we want to investigate how our method can be coupled to future, full-hydro numerical simulations to constrain cosmological parameters without requiring the computation of the mass function nor assuming specific scaling laws.


\section*{Acknowledgments}
We acknowledge the anonymous referee for suggestions that improved the discussion of the results.
The authors thank P.~Valageas for useful discussions on the method.
The present study has been partially supported by a grant from the Centre National d'Etudes Spatiales.  TS acknowledges support from the ESA PRODEX Programme ``XMM-LSS", 
from the Belgian Federal Science Policy Office and from the 
Communaut\'e fran\c caise de Belgique - Actions de recherche concert\'ees. 
FP acknowledges support from Grant No. 50 OR 1003 of the
Deutsches Zemtrum f\"ur Luft- und Raumfahrt (DLR) and from the
Transregio Programme TR33 of the Deutsche
Forschungsgemeinschaft (DfG).

\appendix

\section[]{A realistic model for the observed mass and redshift distributions}
\label{massdist}
In this section, we describe our modeling of the derivation of  $\dndzdm$, in quantity solely used for comparison purpose in the present article. In order to realistically introduce the mass error measurements in the Fisher analysis, we carefully reproduce the various steps involved in the mass determination. In this way, we are able to model how the mass accuracy depends on the data, namely, on the number of photons collected by the instrument.

Using the same ingredients as presented in Sect.~\ref{ingredients}, we compute the expected $dn/dz/dT/dL_X$ distribution of the clusters selected passing the C1 selection, which is based on the total count-rate and apparent extent.
We then assume that the observer is able to measure the temperature $T$ from the collected X-ray photons. The accuracy of such a temperature measurement depends mainly on the number of photons but also on the cluster temperature and requires a prior knowledge of the cluster redshift to a relatively good precision. The influence of these factors on the corresponding observable $\dndzdm$ is discussed in Sect.~\ref{measurement_errors}.

\begin{figure*}
\includegraphics{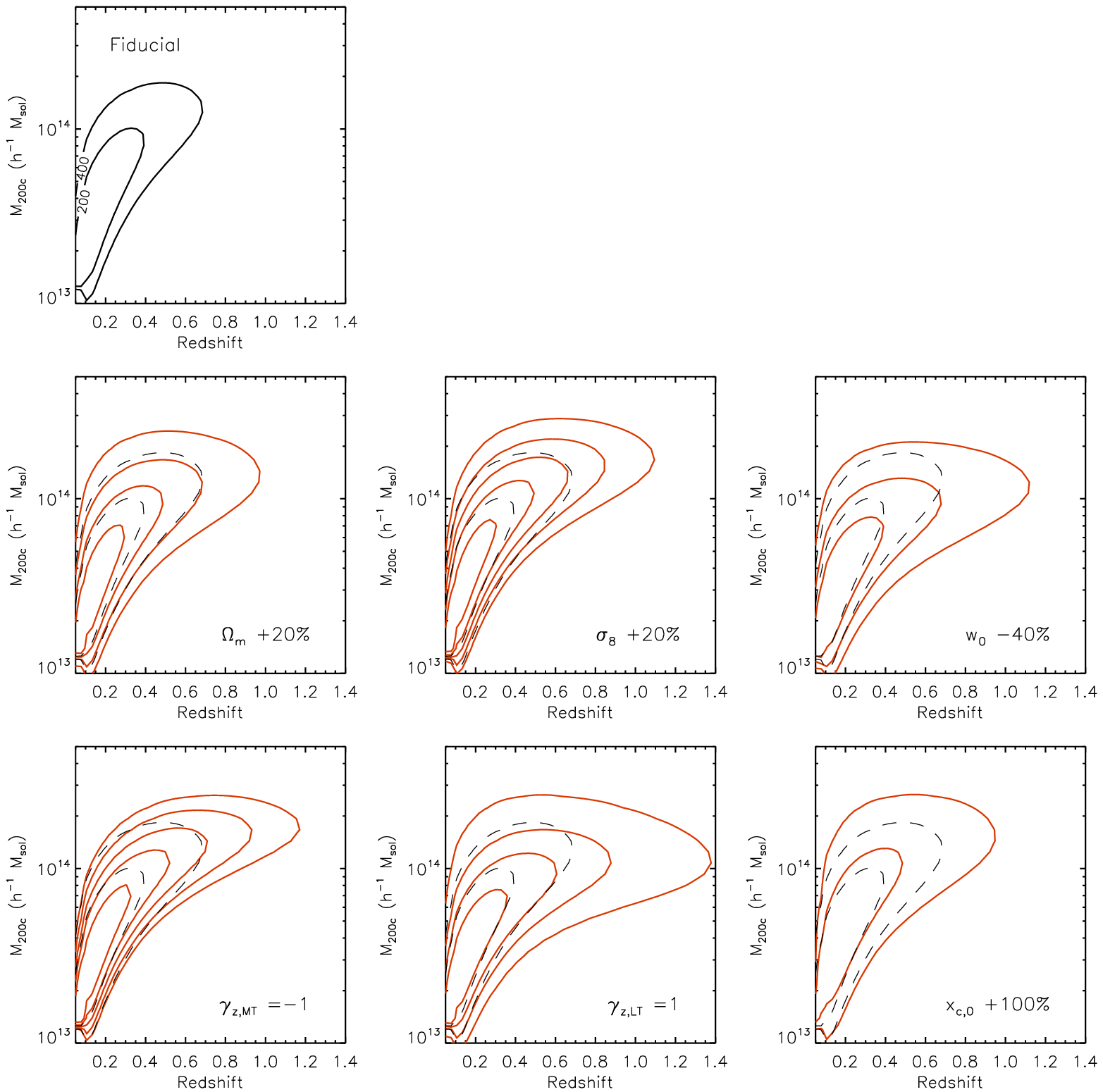}
	\caption{Dependence of the redshift-mass distribution of clusters on the six free model parameters. The fiducial model (Table~\ref{table_parameters}) is represented by the black-dash contours and predicts 570 clusters over 100\,deg$^2$ (10\,ks XMM exposure, C1 selection). The red contours represent the model obtained when one parameter at a time is varied. The corresponding values for each parameter are: 	$\Om$~=~0.30 (+20\%), $\sigma_8$~=~0.94 (+20\%), $w_0$~=~-0.6 (-40 \%), $\zevolmt$~=~-1, $\zevollt$~=~1 and $\xc$~=~0.2 (+100 \%).
	Contour levels stand for the number of clusters enclosed by each curve and are drawn by steps of 200. The differential quantity $\dndzdm$ is constant along each contour.}
	\label{fig_z_m_allpar}
\end{figure*}

The cluster mass is finally obtained by converting $T$ via a mass-temperature relation with parameters chosen in advance. The choice of this ``reference" parameter set is necessary to perform the Fisher analysis described in Sect.~\ref{fisher}.
In the real data analysis, one may also choose to recompute masses with the current parameter values at which the likelihood is estimated (see e.g. \citealt{Vikhlinin:2009p1249, Mantz:2010p1259}).
In practice, we integrate $dn/dz/dT/dL_X$ over $L_X$, then at each $z$ we convert $T$ into $\mdcc$ using eq.~\ref{equ_mt} and a ``reference" parameter set equals to the fiducial model (Table~\ref{table_parameters}). We finally redistribute the result with a constant scatter $\sigma =  \powmt \times \scattmt \sim \sigma_{ln M|T}$ to account for the intrinsic scatter in the scaling law. Thus in all this paper, the quantity $\mdcc$ refers to the mass obtained from the temperature proxy. The shape of the $dn/dz/dM$ distribution for the fiducial model is shown on Fig. \ref{fig_z_m_allpar} along with its dependence on the various parameters of the model (to be compared with Fig. \ref{fig_cr_hr_allpar}).

\section[]{Energy ranges for CR-HR diagrams}
\label{spectral_hr}
This Appendix presents the practical considerations that led to the particular choice of the energy bands used in the article.

\begin{figure}
	\includegraphics[width=84mm]{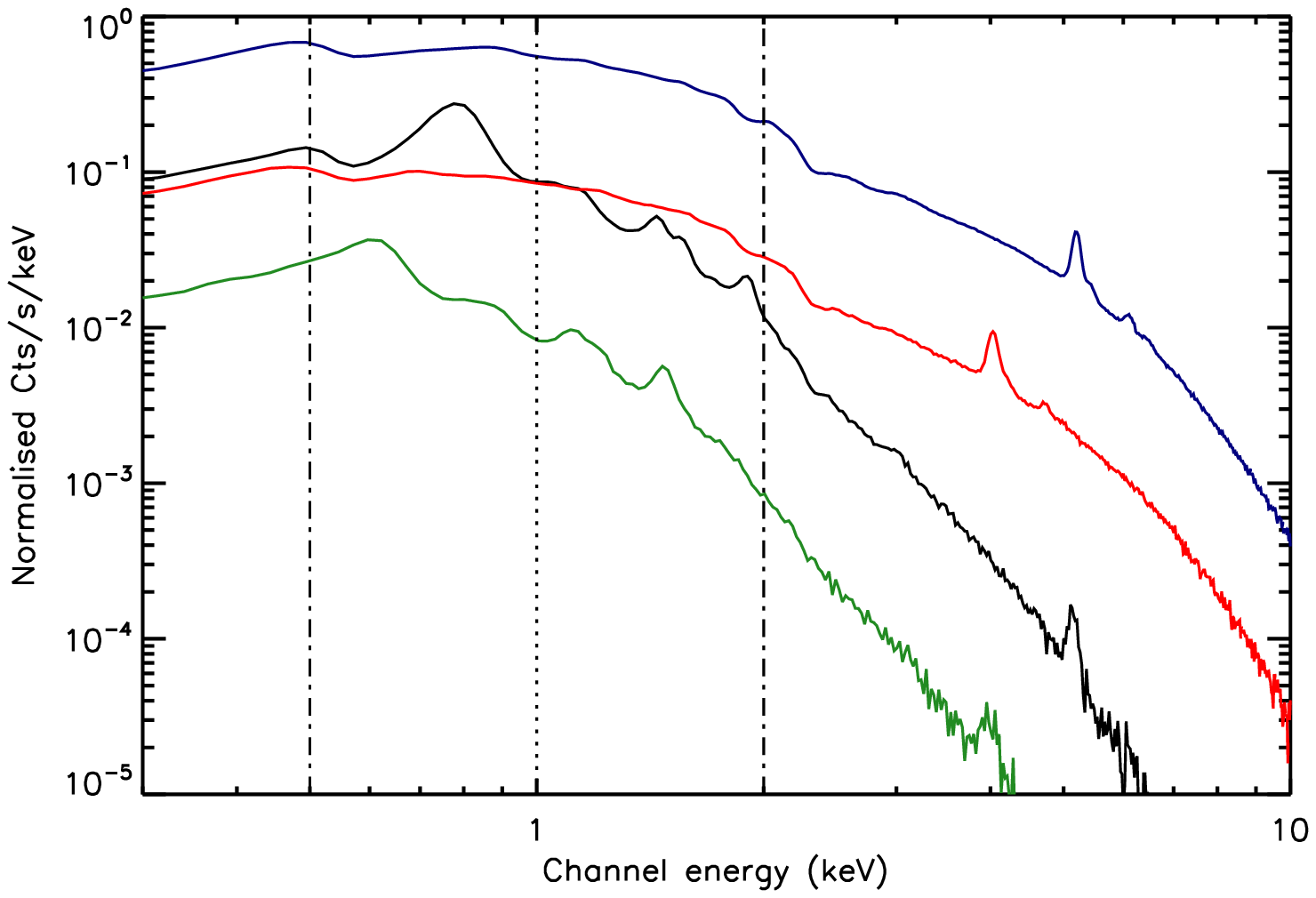}
 \caption{Synthetic APEC spectra convolved by the XMM response, as they would be observed at a very high signal-to-noise level. Vertical lines correspond to 0.5, 1 and 2~keV, i.e. the limits of the 3 energy bands of interest. Each of these spectra is defined by $z$, $T$~(keV) and $L_X$~($10^{44}$~ergs/cm$^2$/s) with the following values: {\it Black: }(0.3, 1.1, 0.5)~; {\it Green: }(0.7, 1.1, 0.5)~; {\it Blue: }(0.3, 4.7, 5.1)~; {\it Red: }(0.7, 4.7, 5.1).}
 \label{fig_spectres} 
\end{figure}

Figure~\ref{fig_spectres} displays four synthetic XMM cluster spectra for typical temperatures and redshifts.
The [0.5-2]~keV band is optimal for cluster detection, given the telescope response and background levels \citep{Scharf:2002p5350}. Moreover, the flux to count-rate conversion in this energy range weakly depends on the temperature for $0.5<T<15$~keV and $0<z<1$. Consequently, the count-rate in [0.5-2]~keV reflects the overall normalization of the cluster X-ray spectrum and is directly related to the cluster bolometric luminosity.

A rough estimate of the cluster spectral shape is the hardness ratio, basically the ratio between two count-rates (or flux measurements) in two energy bands (see e.g. \citealt{Bohringer:2001p1461} for {\it ROSAT} clusters and a different choice of bands).
Because of the particular shape of cluster spectra, the high particle background above 2~keV and the loss of XMM sensitivity, measurement uncertainties in hard bands (typically [2-10]~keV) are high. Thus our low-count clusters (100-1000) are much better characterized at energies below 2~keV. Selected bands must be large enough to minimize the sensitivity to emission features and to maximize the signal-to-noise ratio, but sufficiently narrow to be sensitive to changes in the spectral shape. As shown on Fig.~\ref{fig_spectres}, [1-2] and [0.5-1]~keV appear as good compromises.

We note that a deeper exposure (or higher senstivity) could allow for complementary measurements in the hard part of the spectrum.

\section[]{Constraining power of the cluster redshift number counts}
\label{compar_xxl}

We presented in \citet{Pierre:2011p5484} cosmological forecasts based on the redshift distribution of clusters (and their spatial correlation function) over 50\,deg$^2$. To account for the unknown scaling relations in the sample, we assumed that a mass-luminosity relation can be derived in each of the 20 redshift bins considered. We parametrized the unknown M--L normalizations through 20 parameters $\alpha_i$ (one per redshift bin), further marginalized over when extracting cosmological constraints. We did not put any prior on the normalization of the scaling law nor assumed a functional form for their evolution, but we implicitely supposed that the cluster mass and luminosity can be derived for each individual cluster in the sample (directly from X--ray data or from additional, multi-wavelength, observations).

The present $dn/dz$ analysis differs in that local scaling laws are supposed to be known at a fairly good precision, requiring the call to an external work.
This is rendered by putting stringent priors on the parameters by which they are defined (Table~\ref{table_parameters}). Moreover, the evolution of their normalization with redshift involves two factors on the form $(1+z)^{\zevolmt}$ and $(1+z)^{\zevollt}$. On the other side, only the redshift histogram enters the fitting procedure and there is no need to compute physical properties of clusters at any moment in the analysis.

Despite the difficulty of matching Fisher forecasts obtained by different modeling, we performed a comparison between both approaches. In the former one, we let $\Om$ and $\sigma_8$ free in the Fisher analysis, while priors on the $\alpha_i$'s were set according to \citet{Pierre:2011p5484}, i.e.~assuming a mass accuracy of $\Delta {\rm ln} M = 0.5$ (corresponding to a 10\,ks XMM exposure, C1 cluster selection).
In the current analysis, we put priors on the normalization of the L--T relation as well as on its evolution parameter such that $\sigma(\normlt)=0.02$ and $\sigma(\zevollt)=0.5$ with a correlation of $\rho(\normlt,\zevollt)=-0.9$ in order to mimic the priors on the $\alpha_i$'s that would come from the data itself.
We found $(\Delta \Om \sim 0.03,\, \Delta \sigma_8 \sim 0.06)$ with the $\alpha$-method and $(\Delta \Om \sim 0.02,\, \Delta \sigma_8 \sim 0.07)$ with the present method.
Keeping in mind the difficulty to account for the various parameter degeneracies and the modeling differences between both approaches, we conclude from this comparison that they are consistent with each other.

\section[]{Inventory of the sources of uncertainty}
\label{ERRORS}

In this section we review a number of sources of uncertainty and to which extent they were modeled in the present analysis. Note that most of them are relevant both for the CR-HR approach and the traditional methods based on cluster mass estimates.

	\subsection{Cluster scaling laws}
Fitting simultaneously cosmological and scaling laws parameters as considered here, tends to minimize selection biases as it fully takes into account the sample selection function and provides values of the scaling laws for a full range of cosmological parameters (see \citealt{Mantz:2010p1258} for a recent application).
In particular strong correlations exist between cosmological and scaling laws parameters, whatever the method used: apart from the well-known $\Om$-$\sigma_8$ degeneracy, figures~\ref{fig_crhr_depth_dependence} and~\ref{fig_dndzdm_vs_dzdcrdhr_spectro} show that evolutionnary parameters $\zevolmt$ and $\zevollt$ also correlate strongly with cosmological parameters.

However, our procedure requires prior assumption of a model for the M-T and L-T relations. For this work, we have chosen two power laws (equations~\ref{equ_mt} and~\ref{equ_lt}) with constant scatters, motivated by the observation of individual galaxy clusters (\citealt{Arnaud:1999p398}, \citealt{Arnaud:2005p392}, \citealt{Pratt:2009p322}) and the hydrostatic equilibrium formalism. However, any physically motivated modification of the scaling laws could in principle be parametrized and studied along with the other parameters.
In particular, we did not include any evolution of the scatter in scaling laws, for which observational evidence is weak, nor did we introduce intrinsic correlation between luminosity, temperature and mass (\citealt{Nord:2008p5300}, $\rho_{ltm}$ in \citealt{Mantz:2010p1259}).
For the purpose of comparing different methods, we consider these latter parameters to have a reduced impact over our results.

	\subsection{Measurement errors}
Our models of measurement errors intend to include the main sources of uncertainty arising in real X-ray cluster analyses.
Our relative error on the [0.5-2]~keV measured count-rate amounts to $\sim$~6~\% for a typical 500 counts cluster, roughly consistent with past and current analyses (\citealt{Bohringer:2001p1461}, \citealt{Pacaud:2007p250})
Our assumed measurement error on $ln \mdcc$ for a typical $10^{14}\ h^{-1}\ {\rm M}_{\odot}$ cluster at redshift 0.4 (yielding $\sim~500$~counts with a 10~ks XMM exposure) is $\sim~0.5$, in agreement with \citet{Pierre:2011p5484}.
We note that mass measurement errors should in principle depend on the assumed cosmology and scaling laws. Throughout this analysis, we neglected such variations, as we expect this effect to be negligible relatively to the already high value of the error. This problem does not affect CR-HR diagrams for which no assumption on cosmology is needed to derive measurement errors.

	\subsection{Additional systematics}

		\subsubsection{Halo mass function uncertainties}
Uncertainties in the predicted cosmological mass function is also a source of systematics in real data analyses. In particular, they may arise from the different halo-finders used by different authors to analyse numerical simulations \citep{Knebe:2011p5412}, and amount up to $\sim$~10\%. In this work we used \citet{Tinker:2008p1554} mass function for $\Delta=200$, which is calibrated to roughly 5\% upon numerical simulations, provided the cosmological model is close to $\Lambda$CDM. As we are comparing constraints from different methods based on the same mass function, we expect such uncertainty to have negligible impact over our results. We consider this to be true for any unaccounted-for systematic error occuring before the conversion from halo mass to observables, in particular the conversion between different mass definitions.

		\subsubsection{Profiles and X-ray spatial variations}
Throughout this work we have assumed a very simple isothermal, spherically symetric, $\beta$-model with fixed $\beta$ for the cluster X-ray profiles. This assumption enters the selection function as it is expressed in terms of the apparent core radius $r_c=\xc.\rccc$. It has been widely shown that $\beta$-models do not exactly reproduce the actual complexity of X-ray cluster profiles.
In particular they cannot account for the cool core/non-cool core discrepancy (see e.g. \citealt{Pratt:2002p2256}, \citealt{Cavaliere:2009p570}) which can lead to selection biases \citep{Eckert2011}. More generally, we neglected spatial variations in the cluster X-ray properties and made the somehow strong assumption that a cluster can be described by only three global quantities $z$, $T$ and $L_X$.
We justify this choice by the fact we are considering surveys in which the observed X-ray counts per cluster collected by the detectors are quite low (between 100 and 1000 counts in general) and do not allow for a refined morphological analysis.

		\subsubsection{Total count-rates in wide apertures}
Measured count-rates in the three bands of interest assume that the entire cluster profile can be integrated out to a  large radius independent of the cluster extent, thus neglecting the uncertainty due to background misestimation. There are two ways of accounting for this systematic: either with large simulated samples of realistic cluster observations then correcting for the flux loss (see e.g. \citealt{Bohringer:2001p1461}), or by individually fitting a PSF-convolved model onto the measured profile and integrating it up to large radii (\citealt{Barkhouse:2006p3255}, \citealt{Pacaud:2007p250}, \citealt{Vikhlinin:2009p1249}). In presence of high-quality data, the second option is often preferred, although it is model-dependent.
We are currently investigating how measurements in multiple fixed angular apertures can help in improving CR-HR diagrams. Even if a model will be needed, we expect it to be parametric and "self-calibrated" the same way as we did for $\xc$.

	\subsection{Sample variance}
Throughout this work we neglected the possible correlations between neighboring bins and in particular did not take into account the sample variance in our analysis. Taking it into account would modify the likelihood expressed in Equ.~\ref{equ_likelihood} (see e.g. \citealt{Lima:2004p4562}) by introducing a covariance matrix linking the binned observables to each other. The net effect of cosmic variance is to lower the constraints on the cosmological parameters, but its expression depends on the exact shape of the window function. In this work we only consider the total surveyed area (100~sq.~deg.) without specifying a survey geometry. We leave this study for future work, but we expect the effect of cosmic variance to have the same impact over each of our observables as they all derive from the same primordial quantity (the distribution in mass and redshift).

\bsp

\label{lastpage}

\end{document}